\documentclass{aa}     
\usepackage{graphics}

\def\cc{\,{\rm cm^{-3}}}
\def\cm2{\,{\rm cm^{-2}}}
\def\kms{\,{\rm {km\,s^{-1}}}}
\def\kkms{\,{\rm {K\,km s^{-1}}}}
\def\co{\,{\rm ^{12}CO}}
\def\13co{\,{\rm ^{13}CO}}
\def\h2{\,{\rm H_{2}}}
\def\Msun{\rm M_{\odot}}

\def\aua{{\rm A\&A} }
\def\auas{{\rm A\&AS} }

\def\apj{{\rm ApJ} }
\def\aj{{\rm AJ} }
\def\apjs{{\rm ApJS} }

\def\apjl{{\rm ApJL} }
\def\araa{{\rm ARAA} }
\def\mnras{{\rm MNRAS} }
\def\pasj{{\rm PASJ} }

\begin{document}

\title{CI and CO in nearby galaxy centers} 

\subtitle{The star-burst galaxies NGC~278, NGC~660, NGC~3628, NGC~4631,
  and NGC~4666}
  
\author{F.P. Israel \inst{1}       }
 
   \offprints{F.P. Israel}
 
  \institute{Sterrewacht Leiden, Leiden University, P.O. Box 9513, 2300 RA 
             Leiden, The Netherlands}
 
\date{Received ????; accepted ????}
 
\abstract{} {We study the physical properties and mass of molecular
  gas in the central regions of galaxies with active nuclei.}  {Maps
  and measurements of the $J$=1--0, $J$=2--1, $J$=3--2, $J$=4--3
  $\co$, the $J$=1--0, $J$=2--1 and $J$=3--2 $\13co$ lines in the
  central arcminute squared of NGC~278, NGC~660, NGC~3628, NGC~4631,
  and NGC~4666, as well as 492 GHz [CI] maps in three of these are used
  to model the molecular gas.}  {All five objects contain bright CO
  emission in the central regions. Clear central concentrations were
  found in NGC~660, NGC~3628, and NGC~4666, but not in the weakest CO
  emitters NGC~278 and NGC~4631. In all cases, the observed lines
  could be modeled only with at least two distinct gas components. The
  physical condition of the molecular gas is found to differ from
  galaxy to galaxy. Relatively tenuous (density 100-1000 $\cc$) and
  high kinetic temperature (100-150 K) gas occur in all galaxies,
  except perhaps NGC~3628, mixed with cooler (10-30 K) and denser
  ($0.3-1.0\,\times\,10^{4}\,\cc$) gas. The CO-to-$\h2$ conversion
  factor $X$ is typically an order of magnitude less than the
  `standard' value in the Solar Neighborhood in all galaxy centers.
  The molecular gas is constrained within radii between 0.6 and 1.5
  kpc from the nuclei. Within these radii, $\h2$ masses are typically
  $0.6-1.5\,\times\,10^{8}$ M$_{\odot}$, which corresponds to no more
  than a few per cent of the dynamical mass in the same region.}  {} {
   }

\titlerunning{CI and CO in star-burst spirals}

\maketitle

\section{Introduction}

Molecular hydrogen ($\h2$) is a major constituent of the interstellar
medium in late-type galaxies. Dense clouds of molecular gas are
commonly found in the arms of spiral galaxies, but are frequently also
concentrated in the inner few kiloparsecs.  These concentrations of
gas play an important role in the evolution of galaxy centers.  They
provide the material for the occurrence of inner galaxy star-bursts and
the accretion of super-massive black holes in the nucleus.  It is
therefore important to determine their characteristics (density,
temperature, excitation) and especially their mass but molecular
hydrogen itself is very hard to observe directly.  Instead, its
properties are usually inferred from observations of tracer elements,
notably the relatively abundant and easily excited CO molecule.
Unfortunately, the CO emitting gas is not in LTE, and the most
commonly observed $\co$ lines are optically thick.  We have to observe
CO in various transitions to obtain reliable physical results and also
in an optically thin isotope such as $\13co$ in order to break the
temperature-density degeneracy that plagues $\co$ intensities.  The
measured molecular (and atomic) line intensities are the essential
input for modeling of the physical state of the gas.

We have observed a sample of nearby spiral galaxy centers in various
CO transitions and in the 492 GHz $^{3}$P$_{1}$--$^{3}$P$_{0}$ [CI]
transition.  These galaxies were selected to be bright at infrared
wavelengths, and more specifically to have IRAS flux densities
f$_{12\mu m}\,\geq\,1.0$ Jy.  The results for a dozen galaxies from
this sample have already been published. These are NGC~253 (Israel et
al. 1995), NGC~7331 (Israel $\&$ Baas 1999), NGC~6946 and M~83 =
NGC~5236 (Israel $\&$ Baas 2001 - Paper I), IC~342 and Maffei 2
(Israel $\&$ Baas 2003 - Paper II), M~51 = NGC~5194 (Israel et
al. 2006 - Paper III), NGC~1068, NGC~2146, NGC~3079, NGC~4826, and
NGC~7469 (Israel 2009 - Paper IV).  In this paper, we present the
results obtained for five bright galaxies undergoing intense star
formation (star-burst) in their inner parts. We have summarized basic
observational parameters of these galaxies in Table\,\ref{galparm}.

\begin{table*}
\caption[]{Galaxy parameters}
\begin{center}
\begin{tabular}{llllll}
\hline
\noalign{\smallskip}
                   & NGC~278		       & NGC~660                  & NGC~3628			& NGC~4631                & NGC~4666 \\
\noalign{\smallskip}
\hline
\noalign{\smallskip}
Type$^{a}$     	      & SAB(rs)b                 & SB(s)a pec; HII/Liner    & SAb pec sp  Liner           & SB(s)d                 & SABc: Liner            \\
R.A. (B1950)$^{a}$    & 00$^{h}$49$^{m}$14.6$^{s}$& 01$^{h}$40$^{m}$21.7$^{s}$& $11^{h}17^{m}40.3^{s}$ & $12^{h}39^{m}41.9^{s}$ & $12^{h}42^{m}34.8^{s}$ \\
Decl.(B1950)$^{a}$    &+47$^{\circ}$16$'$44$''$   &+13$^{\circ}$23$'$37$''$   & $+13^{\circ}51'49''$  & $+32^{\circ}48'56''$    & $-00{^\circ}11'19''$   \\
R.A. (J2000)$^{a}$    & 00$^{h}$52$^{m}$04.3$^{s}$& 01$^{h}$43$^{m}$02.4$^{s}$& $11^{h}20^{m}17.0^{s}$ & $12^{h}42^{m}08.0^{s}$ & $12^{h}45^{m}08.6^{s}$ \\
Decl.(J2000)$^{a}$    &+47$^{\circ}$33$'$02$''$   &+13$^{\circ}$38$'$42$''$   & $+13^{\circ}35'23''$  & $+32^{\circ}32'29''$    & $-00^{\circ}27'43''$\\
$V_{\rm LSR}^{b}$      & +630 $\kms$ 	         & +841 $\kms$              & +838 $\kms$ 	   & +618 $\kms$             & +1523 $\kms$\\
Inclination $i^{b}$   & 28$^{\circ}$              & 70$^{\circ}$              & 87$^{\circ}$	   & 86$^{\circ}$            & 70$^{\circ}$ \\
Position angle $P^{b}$&116$^{\circ}$              & 45$^{\circ}$              & 102$^{\circ}$	   & 86$^{\circ}$            & 225$^{\circ}$\\
Distance $D^{a, c}$   & 12.1 Mpc 	         & 13.0 Mpc                 & 11.9 Mpc 		   &  7.7 Mpc                & 22.6 Mpc\\
Scale                 & 58 pc/$''$ 	         & 63 pc/$''$               & 58 pc/$''$   	   & 37 pc/$''$              & 109 pc/$''$\\
\noalign{\smallskip}
\hline
\end{tabular}
\end{center}
Notes to Table 1: $^{a}$ NED $^{b}$ Nishiyama \& Nakai. (2000);
van Driel et al. (1995); Irwin \& Sofue (1996); Rand (2000); Walter,
Dahlem \& Lisenfeld (2004) $^{c}$ Moustakas $\&$ Kennicutt (2006)
\label{galparm}
\end{table*}

\begin{table*}
\caption[]{$\co$ observations Log}
\begin{flushleft}
\begin{tabular}{lccccccccc}
\hline
\noalign{\smallskip}
Galaxy & Date    & $T_{\rm sys}$ & Beam & $\eta _{\rm mb}$ & t(int) & \multicolumn{4}{c}{Map Parameters} \\
       &         &              & Size & & & No. & Size & Spacing & P.A. \\
       & 	 & (K)	        & ($\arcsec$) & & (sec) & pnts & ($\arcsec$) & ($\arcsec$) & ($^{\circ}$)	\\
\noalign{\smallskip}
\hline 
\noalign{\smallskip}
$J$=1-0 (115 GHz)\\
NGC~278 & 05Nov &  375 	       &22    & 0.74 & 2400 &  1 & -- & -- & -- \\
NGC~660 & 05Feb &  233 	       &\vline&\vline&  600 &  1 & -- & -- & -- \\
NGC~3628& 05Oct &  305         &\vline&\vline& 1200 &  1 & -- & -- & -- \\
NGC~4631& 06Jul &  220         &\vline&\vline&  780 &  1 & -- & -- & -- \\
NGC~4666& 06Jul &  293         &\vline&\vline&  840 &  1 & -- & -- & -- \\
\noalign{\smallskip}
\hline 
\noalign{\smallskip}
$J$=2-1 (230 GHz)\\
NGC~278 & 05Nov &  554         &12& 0.53 &  960  & 1 & -- & -- & -- \\
  & 91Nov/01Sep &  950/315     &21& 0.69 &600/300& 36& 60$\times$40 & 10 &  0\\
NGC~660 & 05Feb &  250         &12& 0.53 &  360  & 1 & -- & -- & -- \\
    91Sep/02Jan & 1040/602     &21& 0.69 &300/600& 30& 40$\times$70 & 10 & 51\\
NGC~3628& 05Oct &  357         &12& 0.53 & 1200  &  1& -- & -- & -- \\
        & 89Feb & 1335         &21& 0.72 & 1200  & 29& 70$\times$50 & 10 &105\\
NGC~4631& 06Jul &  343         &12& 0.53 &  780  &  1& -- & -- & -- \\
  & 91Apr/01May & 772/302      &21& 0.70 &400/300& 53&120$\times$40 & 10 & 84\\
NGC~4666& 06Jul &  224         &12& 0.53 &  840  & 1 & -- & -- & -- \\
    & 00Dec/01Nov & 1104/325   &21& 0.69 &300/480& 70& 60$\times$130& 10 &  0\\
\noalign{\smallskip}
\hline 
\noalign{\smallskip}
$J$=3-2 (345 GHz)\\
NGC~278 & 00Jul/01Jan&625/440  &14    & 0.63 &240/480 & 45& 50$\times$50 &  8 &  0\\
NGC~660 & 95Jul/00Jul&1039/701 &\vline& 0.61 &960/1080& 29& 30$\times$70 & 10 & 51\\
NGC~3628&94Apr/94Nov & 534/870 &\vline& 0.58 &240/400 & 50&120$\times$40 &  8 &105\\
NGC~4631& 01Nov      &  485    &\vline& 0.63 &  240   & 47& 80$\times$40 &  8 & 84\\
NGC~4666&00Feb/00Jul & 517/441 &\vline& 0.63 &900/2700& 48& 70$\times$70 &  6 &  0\\
\noalign{\smallskip}
\hline 
\noalign{\smallskip}
$J$=4-3 (461 GHz)\\
NGC~278 & 01Jan & 1370         &11    & 0.53 &  600 & 11 & 12$\times$18 &  6 &  0\\
NGC~660 & 99Aug & 3900         &\vline& 0.53 & 2400 &  8 & 18$\times$18 &  6 & 51\\
NGC~3628& 94Apr & 2716         &\vline& 0.50 &  720 & 15 & 36$\times$24 &  6 &105\\
NGC~4631& 96Apr & 1679         &\vline& 0.50 & 2400 &  5 & 24$\times$24 &  6 & 86\\
NGC~4666& 01Nov & 2785         &\vline& 0.52 & 1440 & 13 & 24$\times$24 &  6 &  0\\
\noalign{\smallskip}
\hline
\end{tabular}
\end{flushleft}
\label{gal12colog}
\end{table*}

\begin{figure}[h]
\unitlength1cm
\begin{minipage}[t]{9.cm}
\resizebox{9.cm}{!}{\rotatebox{270}{\includegraphics*{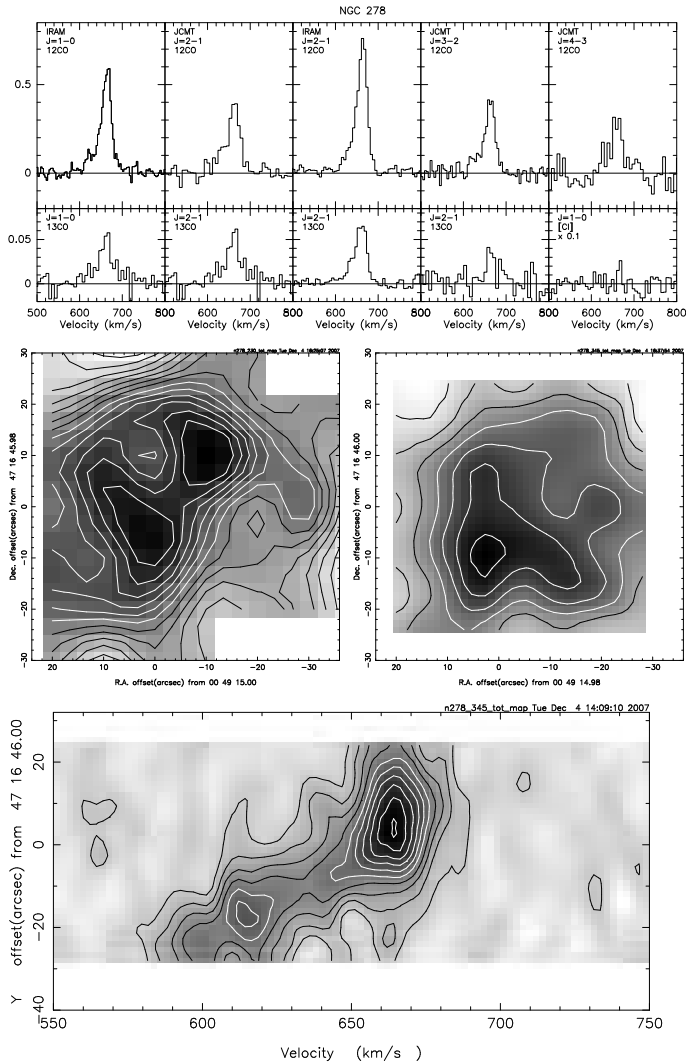}}}
\end{minipage}

\vspace{0.2cm}
\begin{minipage}[t]{9.cm}
\resizebox{4.5cm}{!}{\rotatebox{270}{\includegraphics*{11586f1b.ps}}}
\resizebox{4.5cm}{!}{\rotatebox{270}{\includegraphics*{11586f1c.ps}}}
\end{minipage}

\vspace{0.2cm}
\begin{minipage}[t]{9.cm}
\resizebox{9.0cm}{!}{\rotatebox{270}{\includegraphics*{11586f1d.ps}}}
\end{minipage}
\caption[] {CO in the center of NGC~278.  Top: Observed line
  profiles. Horizontal scale is LSR velocity $V_{\rm LSR}$ in
  km/sec. Vertical scale is main-beam brightness temperature $T_{\rm
    mb}$ in Kelvins, using the efficiencies listed in Table~2. Note
  different scales for $^{12}$CO, $^{13}$CO and [CI] respectively.
  Center: (Left) J=2-1 $^{12}$CO emission integrated over the velocity
  interval 550 - 750 $\kms$. Contours are in steps of 1 $\kkms$; here
  and in all other figures, the first contour is equal to the step
  value. (Right) J=3-2 $^{12}$CO emission integrated over the same
  velocity interval; contour step is 2 $\kkms$.  Bottom:
  J=3-2 $^{12}$CO; position-velocity map in position angle PA =
  45$^{o}$ integrated over a strip 20$''$ wide with contours in steps
  of 0.04 K; northeast is at top. }
\label{n278COmaps}
\end{figure}

\begin{figure}[]
\unitlength1cm
\begin{minipage}[t]{9.cm}
\resizebox{9.cm}{!}{\rotatebox{270}{\includegraphics*{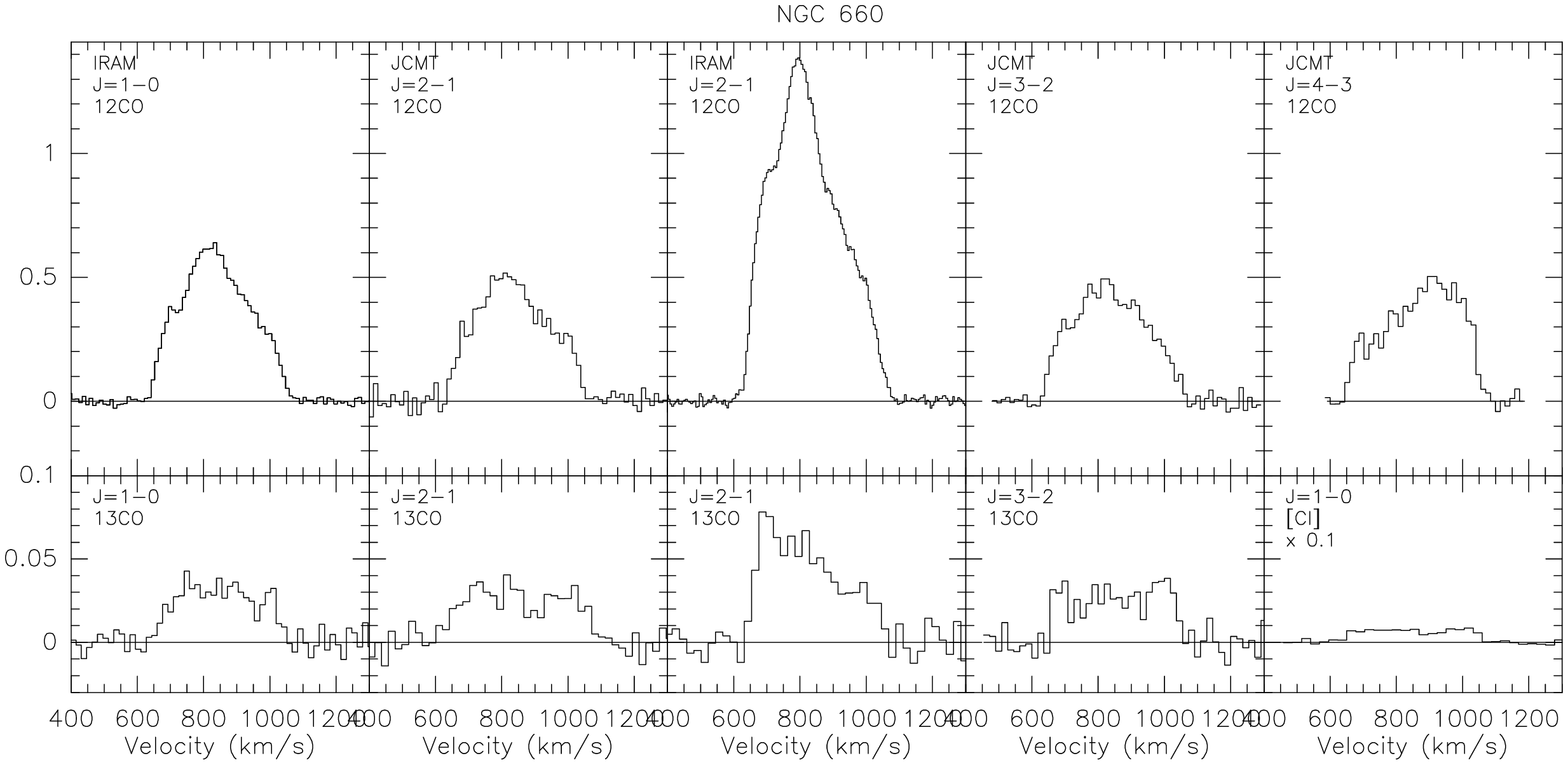}}}
\end{minipage}

\vspace{0.2cm}
\begin{minipage}[t]{9.cm}
\resizebox{4.5cm}{!}{\rotatebox{270}{\includegraphics*{11586f2b.ps}}}
\resizebox{4.5cm}{!}{\rotatebox{270}{\includegraphics*{11586f2c.ps}}}
\end{minipage}

\vspace{0.2cm}
\begin{minipage}[t]{9.cm}
\resizebox{9.0cm}{!}{\rotatebox{270}{\includegraphics*{11586f2d.ps}}}
\end{minipage}
\caption[] {Center of NGC~660, as Figure 1.  Top: Observed line
  profiles. Center: (Left) J=2-1 $^{12}$CO emission integrated over
  the velocity interval 600 - 1100 $\kms$; contour step is 10 $\kkms$;
  (Right) J=3-2 $^{12}$CO emission integrated over the same velocity
  interval; contour step is 15 $\kkms$.  Bottom: J=3-2 $^{12}$CO;
  position-velocity map in position angle PA = 51$^{o}$ integrated
  over a strip 20$''$ wide with contours in steps of 0.04 K; northeast
  is at top. }
\label{n660COmaps}
\end{figure}

\begin{figure}[]
\unitlength1cm
\begin{minipage}[t]{9.cm}
\resizebox{9.cm}{!}{\rotatebox{270}{\includegraphics*{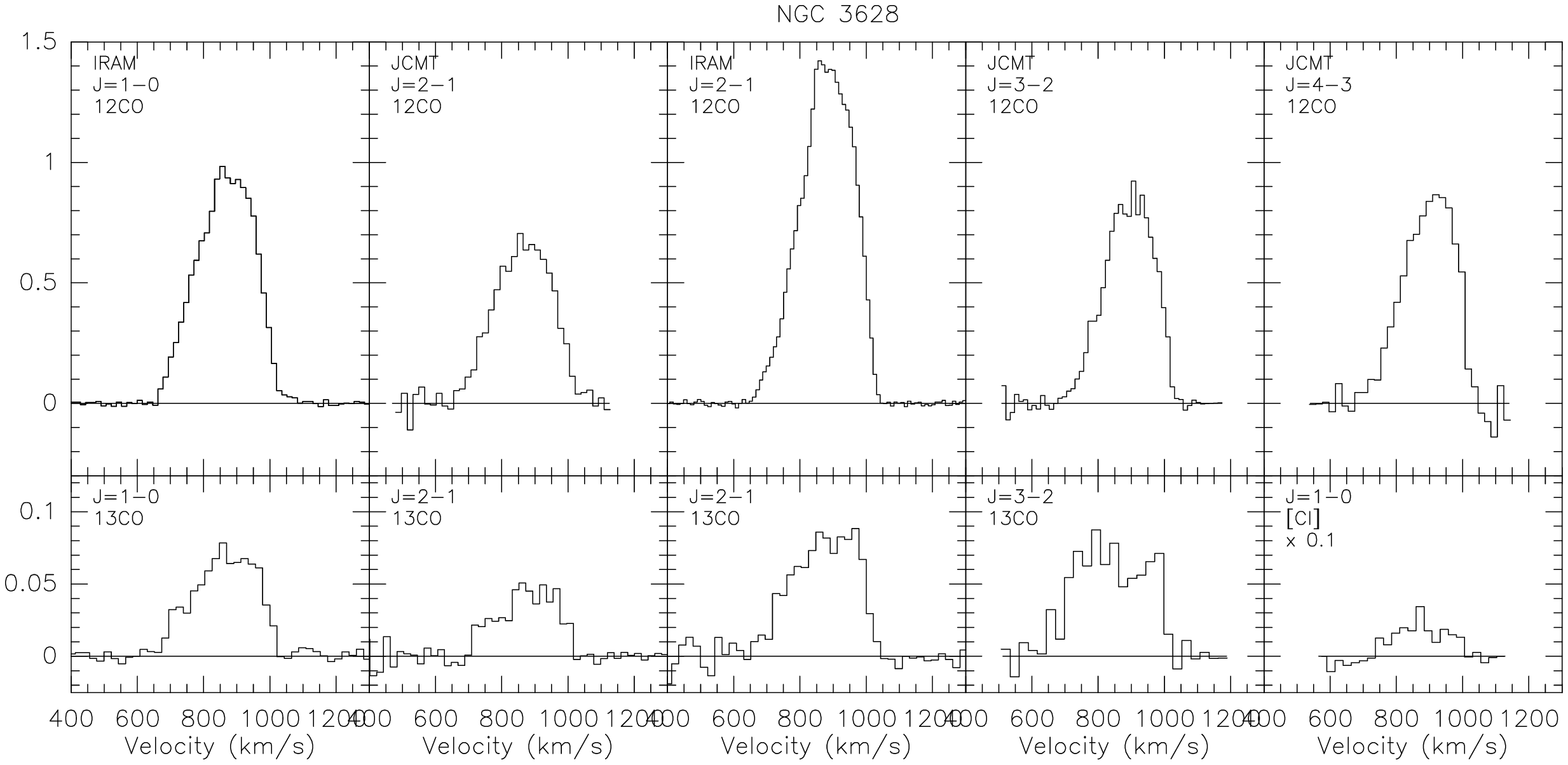}}}
\end{minipage}

\vspace{0.2cm}
\begin{minipage}[t]{9.cm}
\resizebox{4.5cm}{!}{\rotatebox{270}{\includegraphics*{11586f3b.ps}}}
\resizebox{4.5cm}{!}{\rotatebox{270}{\includegraphics*{11586f3c.ps}}}
\end{minipage}

\vspace{0.2cm}
\begin{minipage}[t]{9.cm}
\resizebox{9.cm}{!}{\rotatebox{270}{\includegraphics*{11586f3d.ps}}}
\end{minipage}
\caption[] {Center of NGC~3628, as Figure 1.  Top: Observed line
  profiles.  Center: (Left) J=2-1 $^{12}$CO emission integrated over
  the velocity interval 650 - 1050 $\kms$; contour step is 15 $\kkms$;
  (Right) J=3-2 $^{12}$CO emission integrated over the same velocity
  interval; contour step is 20 $\kkms$.  Bottom: J=3-2 $^{12}$CO;
  position-velocity map in position angle PA = 120$^{o}$ integrated
  over a strip 20$''$ wide with contours in steps of 70 mK; east is at
  top. }
\label{n3628COmaps}
\end{figure}

\begin{figure}[]
\unitlength1cm
\begin{minipage}[t]{9.cm}
\resizebox{9.cm}{!}{\rotatebox{270}{\includegraphics*{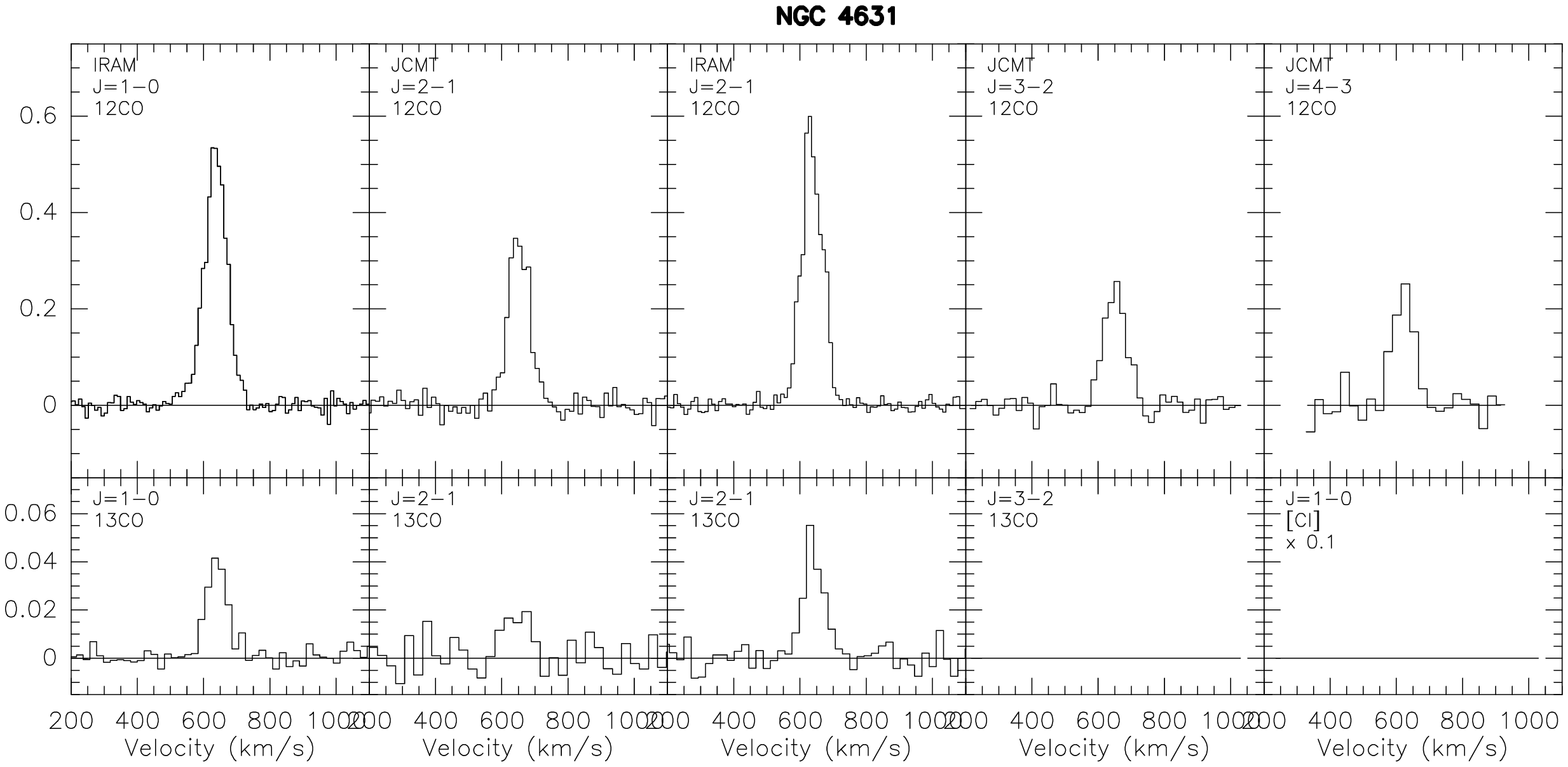}}}
\end{minipage}

\vspace{0.2cm}
\begin{minipage}[t]{9.cm}
\resizebox{4.5cm}{!}{\rotatebox{270}{\includegraphics*{11586f4b.ps}}}
\resizebox{4.5cm}{!}{\rotatebox{270}{\includegraphics*{11586f4c.ps}}}
\end{minipage}

\vspace{0.2cm}
\begin{minipage}[t]{9.cm}
\resizebox{9.cm}{!}{\rotatebox{270}{\includegraphics*{11586f4d.ps}}}
\end{minipage}
\caption[] {Center of NGC~4631, as Figure 1.  Top: Observed line
  profiles. Center: (Left) J=2-1 $^{12}$CO emission integrated over
  the velocity interval 500 - 800 $\kms$; contour step is 4
  $\kkms$; (Right) J=3-2 $^{12}$CO emission integrated over the same
  velocity interval; contour step is 4 $\kkms$.  Bottom:
  J=2-1 $^{12}$CO; position-velocity map in position angle PA = 94
  $^{o}$ integrated over a strip 20$''$ wide with contours in steps of
  40 mK; east is at top. }
\label{n4631COmaps}
\end{figure}

\begin{figure}[]
\unitlength1cm
\begin{minipage}[t]{9.cm}
\resizebox{9.cm}{!}{\rotatebox{270}{\includegraphics*{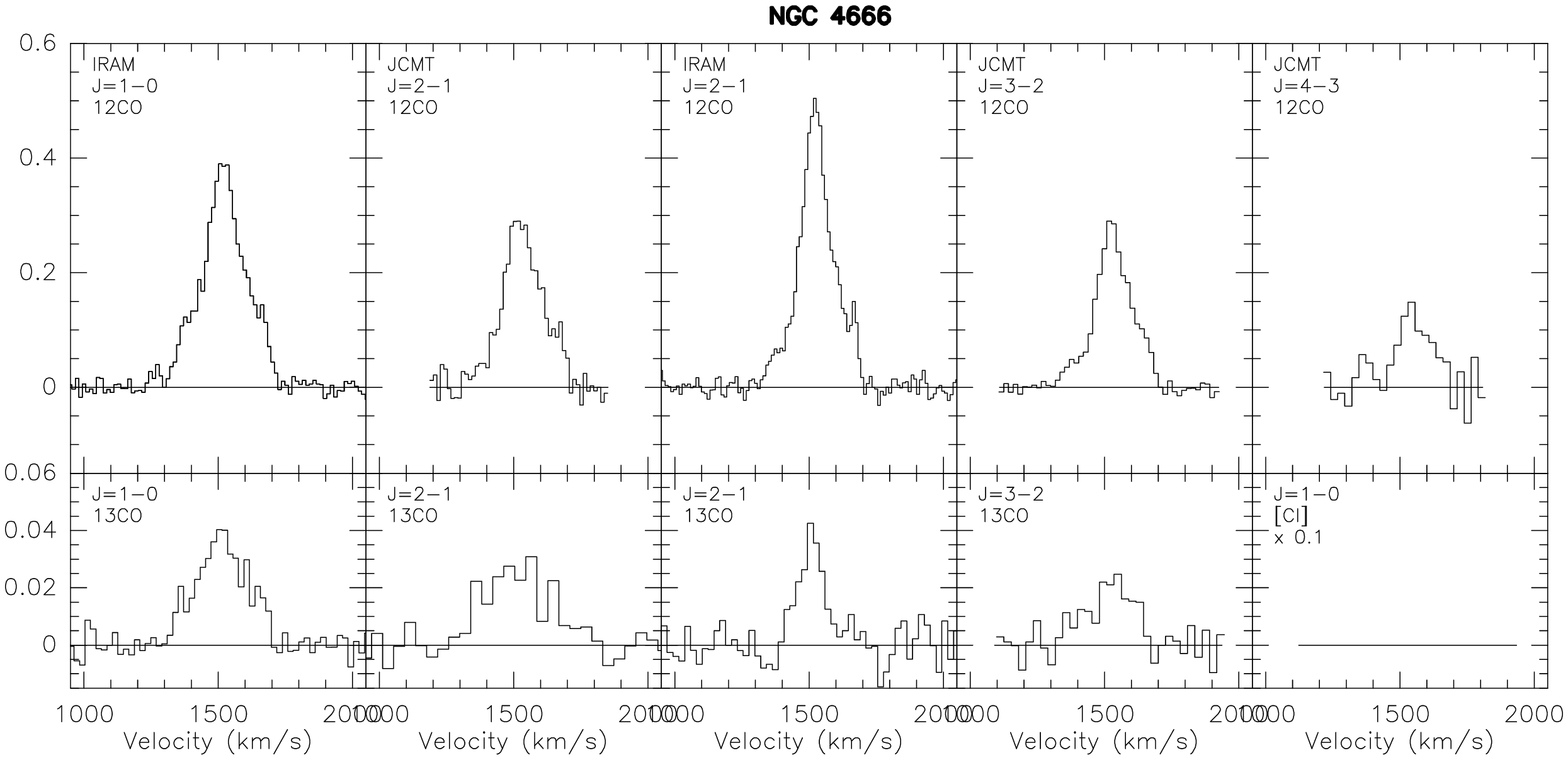}}}
\end{minipage}

\vspace{0.2cm}
\begin{minipage}[t]{9.cm}
\resizebox{4.5cm}{!}{\rotatebox{270}{\includegraphics*{11586f5b.ps}}}
\resizebox{4.5cm}{!}{\rotatebox{270}{\includegraphics*{11586f5c.ps}}}
\end{minipage}

\vspace{0.2cm}
\begin{minipage}[t]{9.cm}
\resizebox{9.cm}{!}{\rotatebox{270}{\includegraphics*{11586f5d.ps}}}
\end{minipage}
\caption[] {Center of NGC~4666, as Figure 1.  Top: Observed line
  profiles. Center: (Left) J=2-1 $^{12}$CO emission integrated over
  the velocity interval 1250 - 1800 $\kms$; contour step is 5 $\kkms$;
  (Right) J=3-2 $^{12}$CO emission integrated over the same velocity
  interval; contour step is 5 $\kkms$.  Bottom: J=3-2 $^{12}$CO;
  position-velocity map in position angle PA = 45$^{o}$ integrated
  over a strip 20$''$ wide with contours in steps of 15 mK; east is at
  top. }
\label{n4666COmaps}
\end{figure}

\begin{table}[h]
\caption[]{$\13co$ and [CI] observations Log}
\begin{flushleft}
\begin{tabular}{llcccccccccc}
\hline
\noalign{\smallskip}
Galaxy &Date&$T_{\rm sys}$ & Beam & $\eta _{\rm mb}$ & t(int) \\
       &    &             &      &                 &        \\
       &    & 	(K)	  & ($\arcsec$) &      	   & (sec)  \\
\noalign{\smallskip}
\hline 
\noalign{\smallskip}
$J$=1-0 (110 GHz)\\
NGC~278 &05Nov&  147 & 23   & 0.75 & 2220 \\
NGC~660 &05Feb&  130 &\vline&\vline&  780 \\
NGC~3628&05Oct&  162 &\vline&\vline& 1200 \\
NGC~4631&06Jul&  150 &\vline&\vline& 1560 \\
NGC~4666&06Jul&  189 &\vline&\vline& 2520 \\
\noalign{\smallskip}
\hline 
\noalign{\smallskip}
$J$=2-1 (220 GHz) \\
NGC~278 &95Jun&  678 &22  & 0.69 & 4730 \\
        &05Nov&  283 & 13 & 0.55 & 2040 \\
NGC~660 &01May&  351 & 22 & 0.69 & 1800 \\
        &05Feb&  259 & 13 & 0.55 &  780 \\
NGC~3628&95Jun&  324 & 22 & 0.69 & 3000 \\
        &05Oct&  212 & 13 & 0.55 & 1320 \\
NGC~4631&01Mar&  444 & 22 & 0.69 & 1800 \\
        &06Jul&  260 & 13 & 0.55 & 1560 \\
NGC~4666&00Dec&  520 & 22 & 0.69 & 3600 \\
        &06Jul&  498 & 13 & 0.55 & 2520 \\
\noalign{\smallskip}
\hline 
\noalign{\smallskip}
$J$=3-2 (330 GHz)\\
NGC~278 &01Jan&  567 & 14   & 0.63 & 6600 \\
NGC~660 &01Jan&  653 &\vline& 0.61 & 4800 \\
NGC~3628&94Apr& 1129 &\vline& 0.58 & 4800 \\
NGC~4631&02jun&  895 &\vline& 0.63 & 4800 \\
NGC~4666&04Feb& 1073 &\vline& 0.63 &36000 \\
\noalign{\smallskip}
\hline
\noalign{\smallskip}
CI $^{3}P_{1}-^{3}P_{0}$\\
J=1-0 (492 GHz)\\
NGC~278 &96Jul      &3640 & 11   & 0.53 & 1800 \\
NGC~660 &96Jul      &3063 &\vline& 0.50 & 1800 \\
NGC~3628$^{a}$&94Nov&3421 &\vline& 0.50 & 1440 \\
\noalign{\smallskip}
\hline
\end{tabular}
\end{flushleft}
\label{gal13colog}
Note {a}. small map of 8 points, size $18''\times18''$, spacing $6''$, in
PA 114$^{\circ}$. 
\end{table}

Although the appearance of all five galaxies in this paper is
completely dominated by a central star-burst, NGC~4666 and perhaps
NGC~3628 also contain a very modest AGN. NGC~660 shows clear evidence for a
recent merger event, and NGC~278 may also have experienced a recent
(minor) merger. The other three galaxies, NGC~3628, NGC~4631, and
NGC~4666, are all interacting with nearby companions.  All five galaxies
are at similar distances: three of them at about 12 Mpc, one
(NGC~4631) somewhat closer at 8 Mpc, and one (NGC~4666) somewhat more
distant at 23 Mpc. Thus, our best angular resolutions of $10''-14''$
correspond to linear resolutions of typically 600-850 pc.  This is not
sufficient to reveal appreciable structural detail in the central CO
distributions; millimeter array observations are required for that.
However, the multi-transition observations of $\co$ and $\13co$
presented here allow us to characterize the overall physical condition
of the central molecular gas in ways not possible otherwise.

\section{Sample galaxies}

\subsection{NGC~278 = DDO~9}

This is an isolated, relatively small and almost face-on galaxy with a
bright nucleus and multiple, spiral arms.  Its optical size is about
$2'$ corresponding to a diameter of 7 kpc.  However, in HI its
diameter is five times greater (Knapen et al. 2004).  Its dust-rich
central parts are quite bright not only in the H$\alpha$ line, but
also in the infrared continuum (Dale et al. 2000; Sanders et al. 2003)
and the [CII] line (Malhotra et al. 2001) consistent with intense
star-forming activity (Ho, Filippenko $\&$ Sargent 1995; Eskridge et
al. 2002).  Optical images show a bright, compact nucleus, but there
is no indication for even a low-luminosity AGN (Tzanavaris $\&$
Georgantopoulos 2007).  Although NGC~278 has been included in many
recent surveys, the only specific study of this galaxy is the one by
Knapen et al. (2004), who concluded that its appearance most likely
reflects a recent minor merger.  Radio mapping yields a total flux
density $S_{\rm 1.4GHz}$ = 0.13 Jy, a source extent of
$48''\times42''$ and a non-thermal spectral index $\alpha$ = -0.66
(Condon, 1983; Condon et al. 1996; $S_{\nu}\propto\nu^{\alpha}$).  The
high-resolution 1.5 GHz map by Condon (1983) shows diffuse emission,
various discrete features following the spiral arms, but no radio
counterpart to the nucleus.  NGC~278 was detected in the $J$=1--0 and
$J$=2--1 $\co$ transitions by Braine et al. (1993; IRAM, $23''$ and
$12''$) and in the former by Elfhag et al. (1996; Onsala, $33''$) and
Young et al. (1995; FCRAO) whose $45''$ beam was barely sufficient to
resolve the galaxy.  $J$=3--2 $\co$ maps were published by Wielebinski
et al. (1999) and again by Dumke et al. (2001; HHT, $24''$) who
briefly discussed the CO properties of the galaxy.

\subsection{NGC~660}

NGC~660 is a remarkable galaxy seen mostly edge-on but exhibiting
clear merger signs such as out-of-the-plane (`polar ring') tidal
features and two intersecting dust lanes.  It is one of the two major
members of the NGC~628 Group.  Radio continuum maps (Condon, 1980,
Condon et al. 1982; Condon et al. 1990) reveal an extended radio
source with two distinct peaks, separated by 3'' (projected linear
separation 190 pc).  The apparent dimensions of the radio source vary
with beam-size, but a size of $25''\times12''$ ($1500\times800$ pc) is
a reasonable approximation. Flux densities are S$_{1.5GHz}$ = 374 mJy,
$S_{4.9GHz}$ = 156 mJy for the extended source total whereas the peaks
have $S_{4.9GHz}$ = 53 and 41 mJy respectively.  The radio continuum
spectrum is dominated by non-thermal emission but the relatively
shallow spectrum ($\alpha$ = -0.6) suggests a non-negligible amount of
thermal emission from this LINER galaxy.  Indeed, high-density ionized
gas is implied by the H92$\alpha$ radio recombination line emission
detected by Phookun et al. (1998).  High-frequency (15 GHz) radio
continuum maps reveal several compact continuum sources most likely
large complexes of high-density star-forming regions and a compact
nuclear source is lacking at radio wavelengths (Carral et al., 1990)
nor is there any obvious X-ray counterpart to the nucleus (Dudik et
al. 2005).  A detailed study of the galaxy, including HI and $J$=1-0,
$J$=2-1 $\co$ mapping (Nobeyama, $17''$, IRAM, $12''$) was carried out
by Van Driel et al. (1995).  The authors concluded that NGC~660 is the
result of a merger of two equally massive objects a few billion years
ago. Single dish $J$=1--0 $\co$ measurements towards the center of
NGC~660 have been summarized by Elfhag et al. (1996) and a $J$=1--0
major axis map was presented by Young et al. (1995, FCRAO,
$45''$). Aalto et al. (1995) measured $\co$ and $\13co$ in the two
lower transitions but there appear to be no high-resolution array maps
of the CO emission in this galaxy. Strong absorption towards the
nucleus of NGC~660 has been detected in HI and in OH by Baan et
al. (1992), as well as in H$_{2}$CO (cf. Mangum et al. 2008).  An
outflow was identified by Baan et al. (1992)

\subsection{NGC~3628}

This is an edge-on Sbc galaxy and part of a close group called the Leo
Triplet (Arp~317) which also includes NGC~3627 and NGC~3623.  In this
paper, we adopt a distance of 11.9 Mpc, which is considerably greater
than 6.7 Mpc often found in the literature.  Where appropriate, we
have recalculated dimensions quoted in other papers for the new
distance value.  NGC~3628 is a galaxy well-studied in various
ways. Its highly disturbed HI distribution betrays its intense
interaction with NGC~3627 (Rots 1978; Haynes et al. 1979).  Radio
continuum observations by Hummel (1980), Condon et al. (1982, 1990),
Condon (1987), and Reuter et al. (1991) show an extended central radio
core of size 6$\arcsec$x1$\arcsec$ (350 x 60 pc) and flux density
$S_{1.5GHz}$ = 0.2 Jy with a predominantly non-thermal spectrum
($\alpha$ = -0.86).  OH and HI absorption observations led Schmelz et
al. (1987a, b) to propose the existence of a small, rapidly rotating
circumnuclear disk with an outer radius of 365 pc, i.e nearly twice
the size of the radio continuum central disk.  As in NGC~660,
formaldehyde is seen in absorption against the extended central radio
continuum source suggesting the presence of large masses of
high-density molecular gas (Mangum et al. 2008, and references
therein).  Indeed, the center of NGC~3628 resembles that of NGC~660
also by its complement of compact radio continuum sources (none
coinciding with the nucleus) representing active regions in the
circumnuclear star-burst contributing about $10\%$ to the total radio
continuum emission (Carral et al. 1990), and by its H92$\alpha$ radio
recombination line emission from high-density HII regions
($n_{HII}\,\approx\,10^{4}\cc$ -- Anantharamaiah et al. 1993; Zhao et
al. 1997). The central core of NGC~3628 is also bright in X-rays, and
the galaxy has an extended halo of soft X-ray emission surrounding a
central outflow, apparently a galactic (star-burst-driven) super-wind
bubble (Dahlem et al. 1996, 1998; Strickland et al. 2004a, b; Grimes
et al. 2005; Flohic et al. 2006).  NGC~3628 may (Roberts et al., 2001;
Gonz\'alez-Mart\'in et al. 2006) or may not (Dudik et al. 2005) contain
a low-luminosity AGN.

Single dish $J$=1-0 $\co$ observations have been reported by Rickard
et al. (1982, NRAO, $55''$), Young et al. (1983, FCRAO, 45$''$), and
Braine et al. (1993, IRAM, 22$''$; $J$=2-1, 12$''$, as well).
Higher-resolution ($10\arcsec - 20\arcsec$) $\co$ maps were published
by Boiss\'e et al. (1987, IRAM, $J$=1--0, 22$''$), Israel et
al. (1990, JCMT, $J$=2--1, 21$''$), Reuter et al. (1991; IRAM,
$J$=2--1, 12$''$); major-axis maps of $\co$ and $\13co$ in the
$J$=1--0 transition were presented by Praline et al. (2001, FCRAO,
45$''$).  NGC~3628 was also mapped in the $J$=1-0 $\co$ transition at
high resolutions (3.85$''$) with the Nobeyama array (Irwin $\&$ Sofue
1996).

\subsection{NGC~4631}

NGC~4631 is an edge-on spiral galaxy interacting with another edge-on
spiral neighbor, NGC~4656, and possibly also with the dwarf elliptical
NGC~4627 (cf. Combes 1978). The distribution of HI throughout the
group has been mapped by e.g. Welsher et al (1978) and Rand (1994);
the WSRT providing a $12''\times22''$ beam very similar to the
resolution of our CO measurements.  Both total-intensity and
major-axis-velocity maps show strong warping and corrugation of the
galaxy disk. Notwithstanding its edge-on orientation, many star
formation regions are seen in H$\alpha$ (Crillon $\&$ Monnet 1969) and
UV (Smith et al. 2001), which are evidently located at the nearby edge
of the galaxy. Radio continuum maps (De Bruyn, 1977; Condon 1983;
Goal 1999) as well as (sub)millimeter and far-infrared maps (Braine
et al. 1995; Dumke et al. 2003; Bendo et al. 2006) show the inner $3'$
to be even brighter at these wavelengths, suggesting the occurrence of
a major star-burst within $R\,\leq\,$ 2.5 kpc, with the possibility of
a `ring' at $R\,=\,30''$, or 1.1 kpc.  However, no emission from the
nucleus has been identified. The galaxy does have a non-thermal halo
(Ekers$\&$ Sancisi 1977).

In addition to significant emission from out-of-the plane hydrogen and
dust, the structure of the large diffuse X-ray halo together with the
observed radio structures suggests wind-driven outflow from a mild
star-burst in the central 5 kpc of the disk (Wang et al. 2001,
Strickland 2004a,b). However, the lack of a central concentration in
star formation is remarkable. In this sense, NGC~4631 resembles
NGC~278 more than the other galaxies in this paper.

The major axis of NGC~4631 has been mapped in $J$=1--0 $\co$ and
$\13co$ by Young et al. (1995) and Paglione et al. (2001), both using
the FCRAO telescope with a resolution of 45$''$, and in the $J$=2--1
$\co$ transition by Sofue et al. (1990; IRAM, 13$''$). The galaxy was
fully mapped in the $J$=1--0 and $J$=2--1 transitions of $\co$ by Golla
$\&$ Wielebinski (1994; IRAM, 22$''$ and 12$''$), who also observed
$\13co$ at five different positions in the same transition. They found
that CO emission extends over $7'$ (15 kpc) along the plane of
NGC~4631, but mostly occur within a diameter of $2'$ ( 4.5 kpc),
corresponding to the star-burst region delineated in radio and X-ray
emission. However, CO is not concentrated to the center.  A map of
NGC~4631 in $J$=3--2 $\co$ was published by Dumke et al. (2001), who
also noted the similarities between NGC~278 and NGC~4631.  Finally, the
inner 100$''$ (3700 pc) of the galaxy were mapped at high resolution
in $J$=1--0 $\co$ by Rand (2000; BIMA, $10''\times7''$), who identified
a molecular component in the outflow from the center as well as
extended CO structure in the disk itself.

\subsection{NGC~4666}

The IRAS survey revealed the highly tilted Sc galaxy NGC~4666 to be an
infrared-luminous star-forming galaxy part of a small group of
galaxies also containing NGC~4668 and NGC~4632 (Garc\'ia 1993; see
also Walter et al. 2004). In the last decade or so, the galaxy has
been observed frequently as part of various surveys, but few detailed
studies exist. In the first of these, Dahlem et al. (1997) found that
the central 5.5 kpc (reduced do our adopted distance of 22.6 Mpc) of
NGC~4666 is the host of a star-burst from which a galactic super-wind
flows out into an extended halo. Star-burst, outflow and halo can be
identified in X-rays (Dahlem et al, 1997, 1998; Persic et
al. 2004). In addition to a star-burst, NGC~4666 also hosts a modest
AGN (Persic et al. 2004; Dudik et al. 2005). NGC~4666 thus presents an
appearance similar to that of the much closer NGC~3079 (see Israel
2009 and references therein).  Low resolution (VLA) radio continuum
maps have been presented by Condon (1983), Condon et al. (1990), and
Dahlem et al. 1997) who found integrated flux-densities of about 0.16
Jy at 4.9 GHz and 0.40 Jy at 1.5 GHz. No nuclear (point source) flux
density has been determined so far; from Condon et al. (1990) an upper
limit is $S_{1.5}\leq0.17$ Jy must be assumed. A major-axis map of
NGC~4666 in $J$=1--0 $\co$ was published by Young et al. (1995; FCRAO,
$45''$). Maps in HI and in the $J$=1--0 $\co$ (SEST: $45''$; IRAM
$21''$; OVRO $4''\times3''$) as well as the $J$=2--1 $\co$ (IRAM:
$11''$) transitions were published by Walter et al. 2004).  These
authors found direct evidence for interactions within the group,
probably causing the star-burst in NGC~4666.

\section{Observations and reduction}

The observations described in this paper were carried out with the 15m
James Clerk Maxwell Telescope (JCMT) on Mauna Kea (Hawaii)
\footnote{The James Clerk Maxwell Telescope is operated on a joint
basis between the United Kingdom Particle Physics and Astrophysics
Council (PPARC), the Netherlands Organisation for Scientific Research
(NWO) and the National Research Council of Canada (NRC).} and with the 
IRAM 30m telescope in Pico Veleta (Spain).
\footnote{The IRAM 30m telescope is supported by INSU/CNRS (France), 
MPG (Germany), and IGN (Spain).}

\subsection{JCMT Observations}  

At the epoch of the mapping observations (1991-2001) the absolute
pointing of the telescope was good to about $3''$ r.m.s. as provided
by pointing observations with the JCMT sub-millimeter bolometer.  The
spectra were calibrated in units of antenna temperature $T_{\rm
  A}^{*}$, correcting for sideband gains, atmospheric emission in both
sidebands and telescope efficiency.  Calibration was regularly checked
by observation of a standard line source.  Further observational
details are given in Tables\,\ref{gal12colog} and \ref{gal13colog}.
Most of the observations were carried out with the now defunct
receivers A2, B3i and C2. Observations in 2001 were obtained with the
newer receivers B3 (330/345 GHz) and W/C (461 GHz). Full details on
these receivers can be found at the JCMT website
(http://docs.jach.hawaii.edu/JCMT/HET/GUIDE/).  Up to 1993, we used a
2048-channel AOS back-end covering a band of 500 MHz ($650\kms$ at 230
GHz). After that year, the DAS digital autocorrelator system was used
in bands of 500 and 750 MHz. Integration times (on+off) given in
Tables\,\ref{gal12colog} and \ref{gal13colog} are typical values
appropriate to the maps.  We subtracted second order baselines from
the profiles.  All spectra were scaled to a main-beam brightness
temperature, $T_{\rm mb}$ = $T_{\rm A}^{*}$/$\eta _{\rm mb}$; values
of $\eta _{\rm mb}$ used are listed in Tables\,\ref{gal12colog} and
\ref{gal13colog}.

All maps were made in rectangular grids (parameters given in
Table\,\ref{gal12colog}), rotated to parallel the major axis of the
galaxy mapped.  The maps shown in this paper all have been rotated
back to a regular grid in right ascension and declination.  

\subsubsection{IRAM observations}

Observations of both $\co$ and $\13co$ in the $J$=1--0 and $J$=2--1
transitions at resolutions of $21''$ and $12''$ respectively were
obtained towards the centers of all galaxies with the IRAM 30m
telescope in the period 2005-2007.  We used the IRAM facility
low-noise receivers A and B in both the 3mm and the 1mm bands. For
back-ends we used the 1MHz and 4MHz filter banks, as well as the VESPA
correlator.  Initially, all four transitions were observed
simultaneously, therefore with identical pointings.  After this, the
signal-to-noise ratio of the $\13co$ profiles was increased by
additional simultaneous observation of the two transitions, where we
took care to ensure that the pointing was the same as for the earlier
observations.  Calibration and reduction procedures were standard and
similar to those described in the previous section.  The assumed $\eta
_{\rm mb}$ values are likewise given in Tables\,\ref{gal12colog} and
\ref{gal13colog}. The IRAM observations were especially important not
only because they provide measurements of the 3mm $J$=1--0 transitions
not possible with the JCMT. The IRAM aperture is twice that of the
JCMT, so that the combination of the two data sets provides
$J$=2--1/$J$=1--0 and $J$=3-2/$J$=2--1 ratios in closely-matched beams,

\begin{table*}[t]
\caption[]{Measured central $^{12}$CO and $^{13}$CO line intensities}
\begin{flushleft}
\begin{tabular}{lccccccccccc}
\hline 
\noalign{\smallskip} 
    & Beam 
    & \multicolumn{2}{l}{NGC~278} 
    & \multicolumn{2}{l}{NGC~660} 
    & \multicolumn{2}{l}{NGC~3628} 
    & \multicolumn{2}{l}{NGC~4631} 
    & \multicolumn{2}{l}{NGC~4666}\\ 
& Size 
& $T_{\rm mb}$ & $\int T_{\rm mb}$d$V$ 
& $T_{\rm mb}$ & $\int T_{\rm mb}$d$V$ 
& $T_{\rm mb}$ & $\int T_{\rm mb}$d$V$ 
& $T_{\rm mb}$ & $\int T_{\rm mb}$d$V$ 
& $T_{\rm mb}$ & $\int T_{\rm mb}$d$V$ \\ 
& ($\arcsec$) 
& (mK) & ($\kkms$) 
& (mK) &($\kkms$) 
& (mK) & ($\kkms$) 
& (mK) &($\kkms$) 
& (mK) & ($\kkms$) \\ 
\noalign{\smallskip} 
\hline
\noalign{\smallskip} 
$\co$ \\
$J$=1-0 & 22     & 595 & 21$\pm$2 &  635 & 163$\pm$16 & 980 & 211$\pm$25 & 550 & 45$\pm$6 & 395 & 77$\pm$9 \\ 
$J$=2-1 & 12     & 755 & 26$\pm$3 & 1377 & 352$\pm$35 &1420 & 207$\pm$25 & 600 & 45$\pm$6 & 505 & 74$\pm$9 \\ 
        & 21     & 393 & 16$\pm$2 &  516 & 137$\pm$14 & 710 & 151$\pm$18 & 390 & 26$\pm$3 & 320 & 54$\pm$6 \\ 
        &{\it 43}& --- & 14$\pm$2 &  --- &  62$\pm$8  & --- & 71$\pm$9   & --- & 21$\pm$3 & --- & 27$\pm$3 \\ 
$J$=3-2 & 14     & 414 & 15$\pm$2 &  490 & 164$\pm$19 & 860 & 163$\pm$20 & 280 & 22$\pm$3 & 315 & 51$\pm$6 \\ 
        &{\it 21}& --- & 13$\pm$2 &  --- &  94$\pm$11 & --- & 130$\pm$15 & --- & 17$\pm$2 & --- & 37$\pm$4 \\ 
$J$=4-3 & 11     & 310 & 15$\pm$2 &  565 & 161$\pm$16 & 930 & 176$\pm$21 & 250 & 19$\pm$2 & 150 & 28$\pm$5 \\ 
        &{\it 14}& --- & 11$\pm$1 &  --- & 122$\pm$14 & --- & 140$\pm$17 & --- & 14$\pm$2 & --- & 23$\pm$3 \\ 
        &{\it 21}& --- &  9$\pm$1 &  --- &  92$\pm$11 & --- &  98$\pm$12 & --- & ---      & --- & 16$\pm$2 \\ 
\noalign{\smallskip} 
\hline
\noalign{\smallskip} 
$\13co$\\
$J$=1-0 & 21 &  68 &2.5$\pm$0.3& 34  & 10.4$\pm$1 & 72  &17.1$\pm$2.0& 40  &3.45$\pm$0.38&40& 8.8$\pm$1.0 \\ 
$J$=2-1 & 12 &  64 &2.7$\pm$0.3& 66  & 18.7$\pm$2 & 82  &20.9$\pm$2.5& 60  &3.75$\pm$0.45&40&11.3$\pm$1.3 \\ 
        & 21 &  77 &1.9$\pm$0.3& 44  & 11.4$\pm$1 & 46  &13.6$\pm$1.7& 20  &1.20$\pm$0.14&30& 8.1$\pm$0.8 \\ 
$J$=3-2 & 14 &  55 &1.3$\pm$0.2& 32  & 12.9$\pm$1 & 77  &20.6$\pm$2.5& --- & ---         &25& 4.5$\pm$0.6 \\ 
\noalign{\smallskip}
\hline
\end{tabular}
\end{flushleft}
\label{galintensity}
\end{table*}

\begin{table*}
\begin{center}
\caption[]{\centerline{Adopted velocity-integrated line ratios in galaxy centres}}
\begin{tabular}{lcccccc}
\hline
\noalign{\smallskip}
Transitions          & Beam   & NGC~278       & NGC~660       & NGC~3628      & NGC~4631      & NGC~4666     \\
                     & ($'')$ &               &               &               &               &              \\
\noalign{\smallskip}
\hline
\noalign{\smallskip}
$\co$ \\
(1--0)/(2--1)  & 21     & 0.89$\pm$0.16 & 1.19$\pm$0.17 & 1.40$\pm$0.28 & 1.72$\pm$0.26 & 1.42$\pm$0.21 \\
(2--1)/(2--1)  & 12/21  & 1.58$\pm$0.24 & 2.57$\pm$0.40 & 1.38$\pm$0.28 & 1.68$\pm$0.25 & 1.37$\pm$0.21 \\
(3--2)/(2--1)  & 21     & 0.79$\pm$0.13 & 0.69$\pm$0.11 & 0.62$\pm$0.12 & 0.66$\pm$0.10 & 0.69$\pm$0.10 \\
(3--2)/(2--1)  & 12     & 0.74$\pm$0.13 & 0.66$\pm$0.12 & 0.87$\pm$0.17 & 0.58$\pm$0.09 & 0.80$\pm$0.12 \\
(4--3)/(2--1)  & 21     & 0.56$\pm$0.08 & 0.66$\pm$0.11 & 0.65$\pm$0.13 &     ---       & 0.30$\pm$0.05 \\
(4--3)/(2--1)  & 12     & 0.54$\pm$0.10 & 0.49$\pm$0.09 & 0.74$\pm$0.15 & 0.46$\pm$0.07 & 0.36$\pm$0.05 \\
\noalign{\smallskip}
\hline 
\noalign{\smallskip}
$\co$/$\13co$ \\
(1--0)         & 21     &  8.4$\pm$1.3  & 15.7$\pm$2.0  & 12.2$\pm$1.8  & 13.0$\pm$2.0  &  8.5$\pm$1.3  \\
(2--1)         & 21     &  8.6$\pm$1.4  & 12.2$\pm$1.8  & 11.7$\pm$1.6  & 22.2$\pm$3.3  &  6.7$\pm$1.0  \\
(2--1)         & 12     &  9.6$\pm$1.5  & 18.8$\pm$2.8  & 13.2$\pm$1.9  & 11.4$\pm$1.6  &  6.3$\pm$1.0  \\
(3--2)         & 14     & 11.4$\pm$1.7  & 12.8$\pm$1.9  &  7.9$\pm$1.8  &      ---      & 11.3$\pm$1.7  \\
\noalign{\smallskip}
[CI]/CO(2--1)) & 21     & 0.29$\pm$0.04 & 0.20$\pm$0.05 & 0.28$\pm$0.06 &      ---      &      ---       \\
\noalign{\smallskip}
\hline
\end{tabular}
\end{center}
\label{galratio}
\end{table*}

\subsection{Results}

Central line profiles in all observed transitions are shown in
Figs.\,\ref{n278COmaps} through \ref{n4666COmaps}.  We mapped the
distribution of the molecular gas in the galaxy centers in the
$J$=2--1, $J$=3--2, and $J$=4--3 transitions, typically over an area
of about one arcminute across.  These maps are also shown in the
Figures, as are the major-axis position-velocity diagrams of the
central CO emission.

Measured line intensities towards the center of each galaxy are listed
in Table\,\ref{galintensity}.  This Table contains entries both at
full resolution, and at resolutions corresponding to those of
transitions observed in larger beams.  These entries were extracted
from maps convolved to the resolution quoted. The convolution was
carried out with the map interpolation function of the SPECX data
reduction package\footnote{
http://docs.jach.hawaii.edu/JCMT/cs/005/11/html/node129.html}. This
function interpolates, where necessary, the map intensities by using
adjacent pixels out to a preset distance (usually 2 beams) and
convolved the observed and interpolated data points with a
two-dimensional Gaussian function (`beam') of predetermined half-width.
This half-width is chosen such that the actual observing beam convolved
with the 2D-Gaussian would yield a beam corresponding to the desired
resolution.  Velocities in a position-velocity map can also be
`smoothed' by applying a convolving Gaussian function in a similar
manner.

Finally, we have determined the line intensity ratios of the observed
transitions at the center of the galaxies observed.  They have been
determined independently from the intensities given in
Table\,\ref{galintensity}. Specifically: (a) we used data taken at the
same observing run wherever possible, (b) we determined ratios at
various resolutions convolving the maps to the relevant beam widths,
and (c) we verified ratios by comparing profile shapes in addition to
integrated intensities.  In Table 5, whenever data were insufficient
for convolution, or altogether lacking (e.g. [CI]), no entry is
provided.

\begin{table*}
\caption[]{Radiative transfer model parameters}
\begin{center}
\begin{tabular}{lccccccccccccc}
\hline
\noalign{\smallskip} 
No. & \multicolumn{3}{c}{Component 1} & \multicolumn{3}{c}{Component 2} & Emiss.&\multicolumn{6}{c}{Model Ratios}\\
     & Kin.       & Gas	& Gradient & Kin.   & Gas      & Gradient    & Ratio &\multicolumn{3}{c}{$\co$} &\multicolumn{3}{c}{$\co/\13co$}\\
     & Temp.      & Dens.    &$\textstyle{N(CO)\over dV}$  & Temp.  & Dens.    & $\textstyle{N(CO)\over dV}$ & Comp.1:2 &$\textstyle{(1-0)\over(2-1)}$&$\textstyle{(3-2)\over(2-1}$&$\textstyle{(4-3)\over(2-1)}$ & (1-0) & (2-1) & (3-2) \\
     & $T_{\rm k}$  & $n(\h2$) & 	  & $T_{\rm k}$ & $n(\h2$)    & 1:2 & \multicolumn{6}{c}{}\\
     & (K)     	    & ($\cc$)    & ($\textstyle{\cm2\over\kms}$) & (K) 	& ($\cc$) &($\textstyle{\cm2\over\kms}$) &\\
\noalign{\smallskip}
\hline
\noalign{\smallskip}
\multicolumn{14}{l}{NGC~278  }\\
 1 & 100 &  500 &  10$\times10^{17}$ &  30 &  10000 & 0.3$\times10^{17}$ & 3:17 &0.92&0.79&0.52& 8.0& 8.9&11.0\\
 2 & 150 &  100 &  30$\times10^{17}$ &  30 &  10000 & 0.3$\times10^{17}$ & 1:4  &0.94&0.80&0.53& 8.2& 9.2&11.5\\
\multicolumn{14}{l}{NGC~660  }\\
 3 & 150 &  500 & 0.6$\times10^{17}$ &  20 & 100000 & 0.3$\times10^{17}$ &11:9  &1.17&0.75&0.49&15.8&11.8&12.9\\
 4 & 150 &  100 & 0.6$\times10^{17}$ & 150 &   3000 & 1.0$\times10^{17}$ & 4:1  &1.17&0.72&0.50&16.3&12.3&12.1\\
\multicolumn{14}{l}{NGC~3628 }\\
 5 & 150 & 1000 & 3.0$\times10^{17}$ &  20 &  10000 & 0.3$\times10^{17}$ & 1:9  &1.10&0.74&0.43&12.8&12.2& 8.9\\
 6 &  20 &  100 & 1.0$\times10^{17}$ &  30 &  10000 & 0.6$\times10^{17}$ &17:3  &1.32&0.63&0.47&12.4&12.3& 8.7\\
\multicolumn{14}{l}{NGC~4631 }\\
 7 & 150 & 1000 & 1.0$\times10^{17}$ &  10 &    300 & 0.45$\times10^{17}$& 1:5  &1.52&0.65&0.42&13.5&10.5&13.2\\
\multicolumn{14}{l}{NGC~4666 }\\
 8 & 100 & 1000 & 1.0$\times10^{17}$ &  10 &  10000 & 0.6$\times10^{17}$ & 7:13 &1.32&0.70&0.37& 8.5& 6.7&11.7\\
\noalign{\smallskip}
\hline
\end{tabular}
\end{center}
\label{galmodel}
\end{table*}

\begin{table*}
\caption[]{Beam-averaged model parameters}
\begin{center}
\begin{tabular}{lccccccccc}
\hline
\noalign{\smallskip} 
Galaxy & &\multicolumn{3}{c}{Beam-Averaged Column Densities} & Outer  & Central    & Face-on       & Relative \\
       & &         &        &                                & Radius & $\h2$ Mass & Mass Density  & Mass \\
       & & $N$(CO) & $N$(C) & $N(\h2)$                       & $R$    & $M_{\h2}$   & $\sigma(\h2)$ & Component 1:2 \\
   &   &\multicolumn{2}{c}{($10^{18} \cm2$)}&($10^{21} \cm2$) & (kpc)  & ($10^{8} \Msun$) &($\Msun$/pc$^{-2}$)& \\
\noalign{\smallskip}
\hline
\noalign{\smallskip}
NGC~278  & 1/2 & 0.24& 0.09& 0.6 & --- & --- &  8 & 0.90:0.10 \\
NGC~660  & 3   & 0.33& 0.38& 1.3 & 0.9 & 0.6 &  7 & 0.67:0.33 \\
         & 4   & 0.49& 0.96& 2.7 & 0.9 & 1.2 & 15 & 0.75:0.25 \\
NGC~3628 & 5   & 2.7 & 0.13& 5.2 & 0.6 & 1.5 & ---& 0.25:0.75 \\
         & 6   & 2.1 & 0.65& 5.1 & 0.6 & 1.5 & ---& 0.90:0.10 \\
NGC~4631 & 7   & 0.22& 0.51& 1.4 & --- & --- & ---& 0.20:0.80 \\
NGC~4666 & 8   & 0.30& 0.31& 1.1 & 1.5 & 1.2 &  6 & 0.50:0.50 \\
\noalign{\smallskip}
\hline
\end{tabular}
\end{center}
\label{galresult}
\end{table*}

\section{Analysis}

\subsection{Radiative transfer modelling of CO}

We have modeled the observed $\co$ and $\13co$ line intensities and
ratios with the large-velocity gradient (LVG) radiative transfer
models described by Jansen (1995) and Jansen et al. (1994) -- but see
also Hogerheijde $\&$ van der Tak (2000) and the web page
http://www.strw.leidenuniv.nl/$\,\tilde{}\,$michiel/ratran/.  These
codes provide model line intensities as a function of three input
parameters per molecular gas component: gas kinetic temperature
$T_{\rm k}$, molecular hydrogen density $n(H_{2})$ and the CO column
density per unit velocity $N({\rm CO})$/d$V$. Full details of the
relevant molecular and atomic line data used in these models can be
found in Sch\"oier et al. (2005).  By comparing model to observed line
{\it ratios}, we may identify the physical parameters best describing
the actual conditions at the observed positions.  Beam-averaged
properties are determined by comparing observed and model {\it
  intensities}. In principle, with seven measured line intensities of
two isotopes, properties of a single gas component are overdetermined
as only five independent observables are required.  In practice, this
is mitigated by degeneracies such as occur for $^{12}$CO.  In any
case, we find that fits based on a single-component gas are almost
always incapable of fully matching the data.

However, we usually obtain good fits based on {\it two} gas
components.  The solutions for a two-component gas are slightly
under-determined but we can successfully compensate for this by
introducing additional constraints as described below.  The physical
gas almost certainly has a much larger range of temperatures and
densities but with the present data two components is the
maximum that can be considered. Indeed, adding even one more component
increases the number of free parameters to the point where solutions
are physically meaningless.  Thus, our two-component model gas is only
an approximation of reality but it is decidedly superior to
one-component model solutions.  As long as significant fractions of
the total molecular gas mass are limited to similar segments of
parameter space, they also provide a reasonably realistic model for
the actual state of affairs.  Specifically, our analysis is less
sensitive to the occurrence of gas at very high densities and
temperatures, but such gas is unlikely to contribute significantly to
the total mass.

In order to reduce the number of free parameters, we assumed identical
CO isotopic abundances for both gas components.  Furthermore, in a
small number of star-burst galaxy centers (NGC~253, NGC~4945, M~82,
IC~342, He~2-10), values of $40\pm10$ have been suggested for the
isotopic abundance [$^{12}$CO]/[$^{13}$CO] (Mauersberger $\&$ Henkel
1993; Henkel et al. 1993, 1994, 1998; Bayet et al. 2004), somewhat
higher than the characteristic value of 20--25 for the Milky Way
nuclear region (Wilson \& Rood 1994).  We therefore have adopted an
abundance value of [$^{12}$CO]/[$^{13}$CO] = 40 in our models.  We
identified acceptable fits by searching a grid of model parameter
combinations (10 K $\leq T_{\rm k} \leq $ 150 K, $10^{2} \cc \leq
n(\h2) \leq 10^{5} \cc$, $6 \times 10^{15} \cm2 \leq N(CO)/dV \leq 3
\times 10^{18} \cm2$) for model line ratios matching the observed set,
with the relative contribution of the two components as a free
parameter. The models rarely predict the precise set of line ratios
that is observed.  The $\co$ ratios are generally derived from
observations taken at different times, and require convolution of one
of the lines to the resolution of the other.  In contrast, $\co/\13co$
line ratios are derived from observations done (quasi)simultaneously
and in with identical beams, so that they are much more accurately
determined.  Thus, when confronted with competing sets of model ratios
in which one or more of the individual ratios differs from the
observed ratio beyond the observational uncertainty, we value a good
fit to the $\co$ transition ratios much less than one to the observed
$\co/\13co$ line ratios.  Comparison of the entries in Table 5 and the
last six columns of Table 6 shows that this choice had to be made for
NGC~3628 (in particular for the $J=4-3/J=2-1$ ratio, and also for
NGC~4631 which is difficult to fit with any model set.

Solutions obtained in this way are not unique, but rather define a
limited range of values in distinct regions of parameter space.  For
instance, variations in input parameters may to some extent compensate
one another, producing identical line ratios for somewhat different
combinations of input parameters.  Among all possible solution sets,
we have rejected those in which the denser gas component is also much
hotter than the more tenuous component, because we consider the large
pressure imbalances implied by such solutions physically implausible,
certainly on the kilo-parsec scales observed.

The results of our model fitting procedure are summarized in
Table\,\ref{galmodel}. The Table lists, under the proviso just given,
representative model solutions reproducing the observed line ratios.
In appropriate cases (e.g. N~278, N~660, and N~3628) , two solutions
are given, each representative of a group of solutions with very
similar parameters.  No galaxy observation set yielded more than two
groups of plausible solutions. The Table gives for each of the two
representative gas components the kinetic temperature $T_{\rm kin}$,
the $\h2$ volume density $n{\h2}$, and the CO
gradient$\textstyle{N(CO)\over dV}$.  The relative contribution of the
two components to the observed $J$=2--1 $\co$ emission is also
given. The last six columns give the corresponding model ratios, which
may be compared directly to the observed ratios (in a $21''$ beam) in
Table\,\ref{galratio}.

\subsection{Beam-averaged molecular gas properties}

Chemical models, such as presented those by van Dishoeck $\&$ Black
(1988) and recently updated by Visser, Van Dishoeck, $\&$ Black
(2009), show a strong dependence of the (neutral and ionized) atomic
carbon to carbon monoxide column density ratio, $N({\rm C})/N({\rm
  CO})$, on the total carbon $N_{\rm C}\,=\,N(\rm C\,)+\,N(\rm CO)$
and molecular hydrogen column densities.  With line intensities $I_{rm
  CO}$, $I_{\rm [CI]}$, and $I_{\rm [CII]}$ of all three species
known, minimal assumptions suffice to deduce total carbon and hydrogen
column densities and masses.

Unfortunately, when we have {\it incomplete or no knowledge} of these
C$^{\circ}$ and C$^{+}$ line intensities, as is the case here and
often elsewhere as well, this procedure is useless.  However, lacking
direct and full atomic carbon information, we may still make progress
by using a new element instead. The total [C]/[H] gas-phase carbon
abundance may serve equally well as a constraining factor in addition
to the CO column density resulting from the radiative transfer model.
The chemical models provide the fraction of CO (i.e. the ratio of
carbon monoxide to atomic carbon, whether neutral or ionized) as a
function of $\h2$ column density.  The fraction derived from the
models varies as a function of $\h2$ volume density and the radiation
field.  In all cases we used the $\h2$ density derived from the LVG
modeling.  For the {\it average} radiation field, we assumed
relatively moderate conditions to apply to the objects studied here,
e.g. radiation field intensities typically about an order of magnitude
higher than those in the Solar Neighborhood, or less. At the same
time, the gas phase carbon abundance provides us with the total carbon
column (i.e. the sum of carbon monoxide and atomic carbon) for any H2
column density.  Thus, each value of $N(\h2)$ is associated with a
unique value of $N(\rm C)/N(\rm CO)$ {\it and} a unique value of
$N_{\rm C}\,=\,N(\rm C)\,+\,N(\rm CO)$.  Knowing any one of the three
column densities (e.g. $N(\rm CO)$, the two equations allow us to find
the remaining columns (e.g. $N(\h2)$ and $N(\rm C)$,).  What fraction
of all atomic carbon (C) is ionized (C$^{+}$) or neutral (C$^{\circ}$)
remains undetermined, but that does not concern us here.

Oxygen abundances and gradients of many nearby galaxies have been
compiled by Vila-Costas $\&$ Edmunds (1992), Zaritsky et al. (1994),
and Pilyugin, V\'ilchez $\&$ Contini (2004).  None of the five
galaxies studied here is included in any of these compilations.  The
first two compilations suggest nuclear metalicities many times solar,
but Pilyugin et al.'s more recent work implies nuclear metalicities on
average only 1.5 times solar.  A major uncertainty in these values is
caused by the extrapolation of disk HII region abundances to the very
different galaxy center environment.  We have conservatively assumed a
metalicity twice solar for the galaxy central regions.  The results
published by Garnett et al. (1999) suggest very similar carbon and
oxygen abundances at these metalicities, so that we adopt [C]/[H]
$\approx\,1.0\,\times\,10^{-3}$.  As a significant fraction of carbon
is tied up in dust particles and thus unavailable in the gas-phase, we
have also adopted a fractional correction factor $\delta_{\rm c}$ =
0.27 (see for instance van Dishoeck $\&$ Black 1988), so that $N_{\rm
  H}/N_{\rm C}$ = [2$N(\h2) + N$(HI)]/[$N$(CO) + $N$(CII) + $N$(CI)] =
3700, uncertain by about a factor of two.  In Table\,\ref{galresult}
we present beam-averaged column densities for CO and C (= C$^{\rm
  o}$+C$^{+}$), as well as $\h2$ derived under these assumptions.  As
the observed peak CO intensities are significantly below the model
peak intensities, only a small fraction of the (large) beam surface
area can be filled with emitting material.

Although the analysis in terms of two gas components is superior to
that assuming a single component, it is not fully realistic.  For
instance, the assumption of e.g. {\it identical beam filling factors}
for the various species ($\co, \13co$, C$^{\circ}$, and C$^{+}$) is
not a priori plausible.  However, as argued before ({\it cf} Israel,
2009), the beam-averaged column-density is modified only by the degree
of non-linearity in the response of the model parameters to a change
in filling factor, {\it not} by the magnitude of that change.  A
similar state of affairs exists with respect to {\it model
  degeneracy}.  Even for rather different model fractions of hot or
dense gas, in any particular galaxy the {\it beam-averaged} molecular
hydrogen densities are practically the same.  The uncertainties
introduced by plausibly different average radiation fields are also
relatively minor. As we are observing with a large linear beamsize, we
are not sensitive to the very highly excited or very dense molecular
gas that only occurs on small scales. In fact, the assumed gas-phase
carbon abundance dominates the results.

The derived beam-averaged column-densities (and projected mass
densities) roughly scale inversely proportional to the square root of
the abundance assumed: $N_{av}\,\propto\,{[C]/[H]_{gas}}^{-0.5}$. As
our estimate of the central elemental carbon abundance, and our
estimate of the carbon depletion factor are unlikely to be wrong by
more than a factor of two, the resulting uncertainty in the final
values listed in Table\,\ref{galresult} should be no more than $50\%$.

\section{Discussion}

\subsection{NGC~278}

The CO distribution shown in Fig.\,\ref{n278COmaps} reveals that there
is no concentration of molecular gas towards the nucleus of the
galaxy, as was already suggested by the lower-resolution map of Dumke
et al. (2001).  Instead, a few local peaks in the extended diffuse
emission occur near positions of enhanced radio continuum emission in
the spiral arms (see Condon, 1986). 

\begin{figure}[]
\unitlength1cm
\begin{minipage}[t]{6.cm}
\resizebox{5.cm}{!}{\rotatebox{270}{\includegraphics*{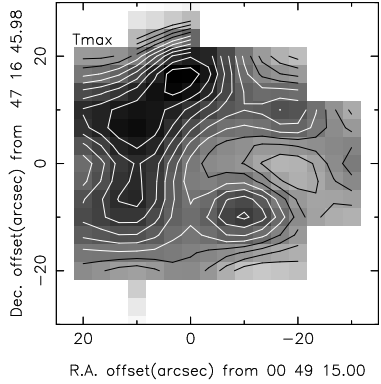}}}
\end{minipage}
\caption[] {Center of NGC~278. Map of peak J=2-1 $^{12}$CO brightness
  temperature $T_{mb}$. Contour step is 30 mK, lowest contour is at
  300 mK.}
\label{N278_Peaks}
\end{figure}

 This is particularly clear in a map of peak rather than integrated
 brightness temperature (see Fig.\,\ref{N278_Peaks}), showing
 unresolved maxima with $T_{mb}$ = 0.5-0.7 K.  These CO and radio
 continuum peaks mark the locations of major star-forming complexes in
 the inner disk.  The molecular gas in the SE arm appears to be
 somewhat warmer than that in the NW arm.  The inner rotation gradient
 in Fig.\,\ref{n278COmaps} is 150 $\kms$/$'$.  Taking into account
 distance and inclination of the galaxy as listed in
 Table\,\ref{galparm}, this corresponds to 90 $\kms$/kpc. The maximum
 rotational velocity width is 110 $\kms$, or 235 $\kms$ corrected for
 inclination.  In the center of the galaxy, the slow decrease of $\co$
 intensity with increasing rotational transition
 (Table\,\ref{galmodel} suggests the presence of a fair amount of warm
 molecular gas. At the same time, the $\13co$/$\co$ isotope intensity
 ratios are not particularly high, implying that reasonably dense gas
 should also be present.  Indeed, a robust result of our modeling
 (Tables 6, 7) is that almost 90 per cent of the molecular gas mass is
 relatively dense ($n_{\h2}\,\approx\,10^{4}\,\cc$) and not very warm
 with a kinetic temperature of about 30 K.  The remaining 10 per cent
 of the molecular mass is much hotter ($T_{\rm k}$ = 100-150 K) and
 much less dense ($n_{\h2}$ =100-500 $\cc$).  Such temperatures and
 densities are not unusual for photon-dominated regions (PDRs)
 associated with the star formation process.  Notwithstanding the
 uncertainty in the hot component parameters, the beam-averaged
 parameters allow for very little variation.  Most of the carbon
 should be in CO, the ratio of C to CO being 0.4.

The implied face-on molecular hydrogen mass density of 8
$\Msun$/pc$^{2}$ is not exceptional.   The relatively low beam-averaged
molecular hydrogen column density $N(\h2)$ = 6$\times$10$^{20}$ $\cm2$
combined with the modest $J$=1-0 $\co$ intensity of 21 $\kkms$ implies
a low value for the CO-to-$\h2$ conversion factor $X$ =
3$\times$10$^{19}$ $\cm2/\kkms$, i.e. typically 6-7 times less than 
the Solar Neighbourhood conversion factor. Again, this is not unusual
and in line with findings for other galaxies ({\it cf} the discussion
in Israel, 2009). 

\subsection{NGC~660}

\begin{figure}
\begin{minipage}[t]{9.cm}
\resizebox{9.cm}{!}{\rotatebox{270}{\includegraphics*{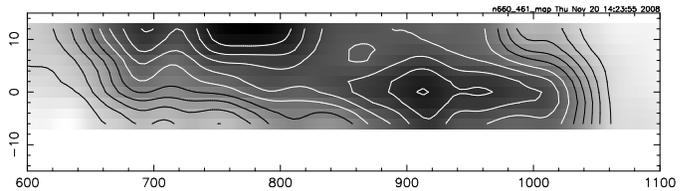}}}
\end{minipage}
\caption[] {Center of NGC~660, $J$=4-3 $^{12}$CO position-velocity map
  in position angle PA = 45$^{o}$ integrated over a strip 14$''$ wide
  with contours in steps of 40 mK; northeast is at top. }
\label{n660more}
\end{figure}

As Fig.\,\ref{n660COmaps} shows, there is a clear CO maximum spanning
the central (deconvolved) $17''$, which means that there is a very
significant molecular gas concentration within a radius $R$ = 500 pc
from the nucleus.  At lower levels of surface brightness, CO emission
is present over the full extent of our maps. Indeed, in the (IRAM 30m)
$\co$ maps published by Van Driel et al. (1995), ever weaker emission
is seen to extend out to radii of $70''$ (4.4 kpc).  The central CO
emission peak in NGC~660 is quite strong (Table 4).  The
major-axis/velocity diagram suggests that this central molecular gas
is in rapid rotation, following a solid-body rotation curve.  The
velocity gradient of rotation is 355 $\kms$/kpc, corrected for
inclination; the full rotational width is 450 $\kms$, again corrected
for inclination. In Figs.\,\ref{n660COmaps} and Fig.\,\ref{n660more},
the very central few arcseconds (corresponding to a radius $R\,\leq$
100 pc) show a pronounced molecular gas minimum.  The size of this
minimum corresponds closely to the separation of the radio peaks
mentioned earlier. The $\co$ ratios in Table 5 suggest that the gas
within a radius of $6''$ ($R\,\leq$ 380 pc) is both cooler (lower
$\co$ (4-3)/(2-1) ratio) and less optically thick (higher $J$=2-1
$\co$/$\13co$ ratio) than that between radii of $11''$ and $6''$ ($R$
= 380 -- 700 pc).  More in general, the observed line ratios may be
explained by either of two models. The corresponding reasonably good
fits (see Table 6) are virtually indistinguishable notwithstanding the
differences in underlying physical conditions. Tables 6 and 7 show
that in the first model, two thirds of the gas mass is hot ($T_{kin}$
= 150 K) and tenuous ($n_{\h2}\,\approx\,500\,\cc$), and the remaining
third is cold ($T_{kin}$ = 20 K) and very dense
($n_{\h2}\,\approx\,10^{5}\,\cc$).  In the second model, essentially
all gas is hot ($T_{kin}$ = 150 K), three quarters is very tenuous
($n_{\h2}\,\approx\,100\,\cc$) and the remaining quarter is only
modestly dense ($n_{\h2}\,\approx\,3000\,\cc$).  Overall, the two
models lead to integrated central masses and projected mass densities
different by a factor two, the latter model yielding the highest
values.  As either of the two solutions yields plausible parameters,
it is not easy to determine which model is the more likely to
apply. We may nevertheless firmly conclude that in the center of
NGC~660 {\it most of the molecular gas is very hot and rather
  tenuous}.  The corresponding CO-to$\h2$ conversion factor in the
center of is $X$ = 0.8-1.6 $\times$10$^{19}$ $\cm2/\kkms$, i.e. 12-25
times {\it lower} than the `standard' Solar Neighborhood value. We
find that the implied central gas mass is 1-2$\%$ of the dynamical
mass in the same volume.

\subsection{NGC~3628}

\begin{figure}[]
\unitlength1cm

\begin{minipage}[t]{9.cm}
\resizebox{4.5cm}{!}{\rotatebox{270}{\includegraphics*{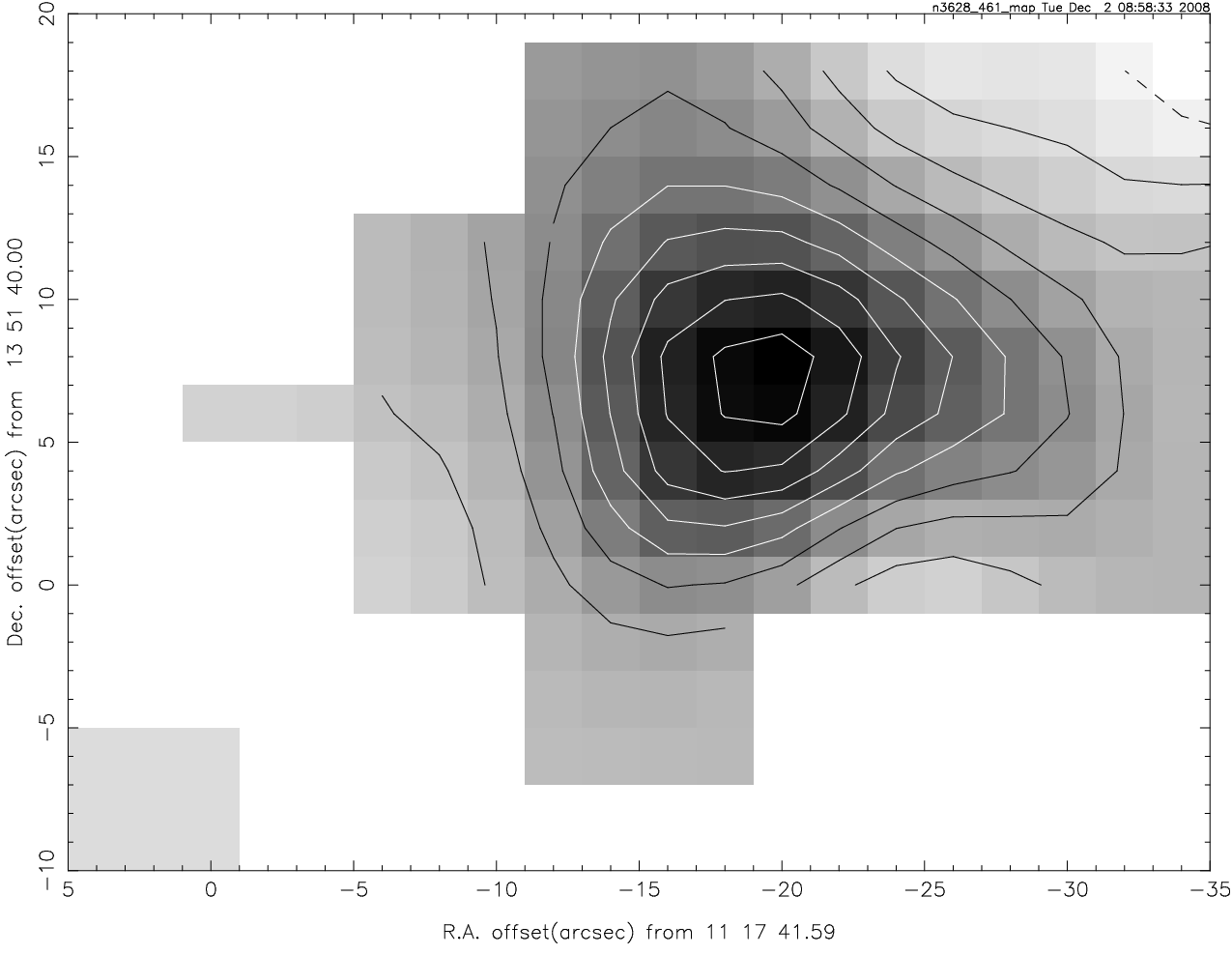}}}
\resizebox{4.5cm}{!}{\rotatebox{270}{\includegraphics*{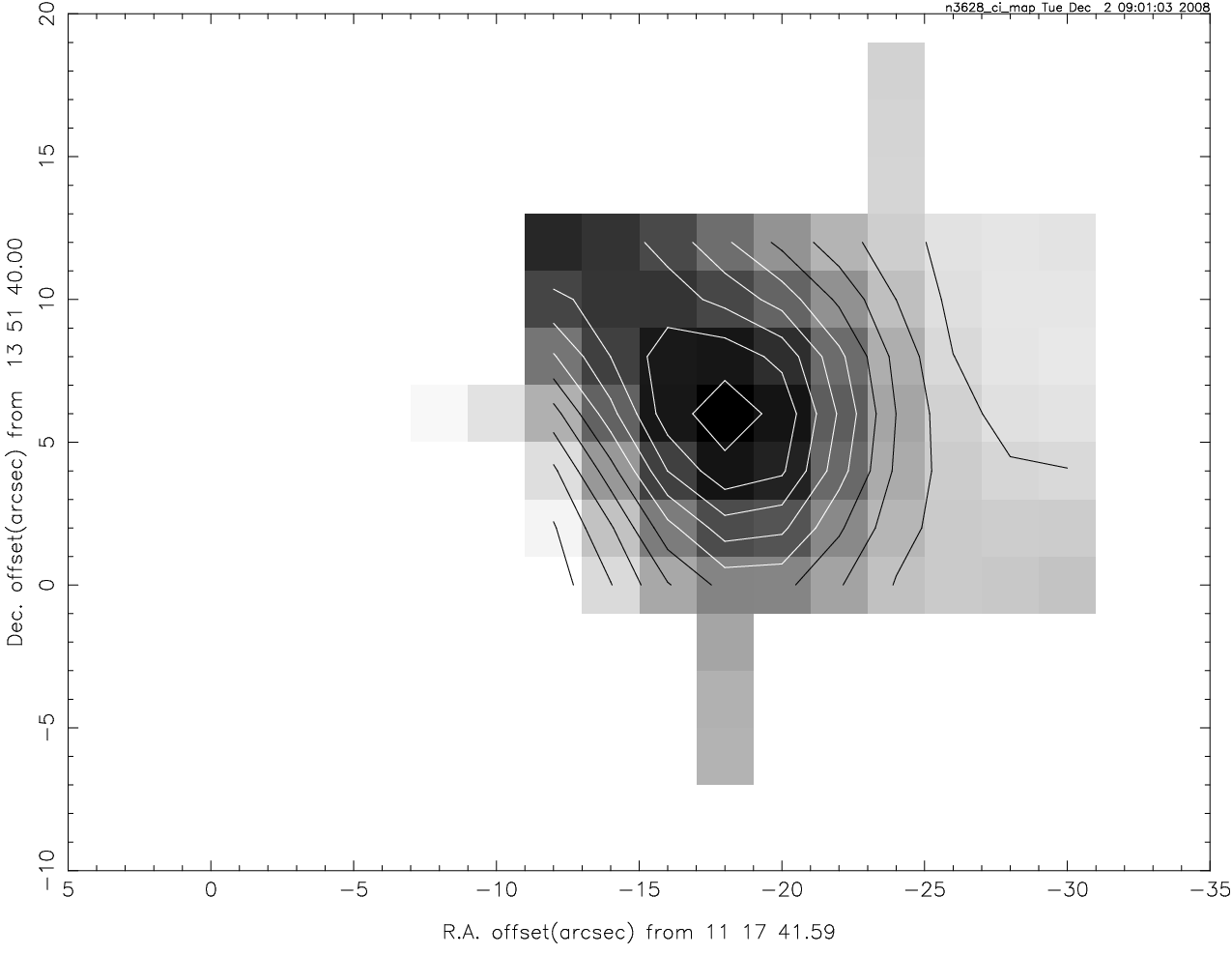}}}
\end{minipage}

\vspace{0.2cm}
\begin{minipage}[t]{9.cm}
\resizebox{9.cm}{!}{\rotatebox{270}{\includegraphics*{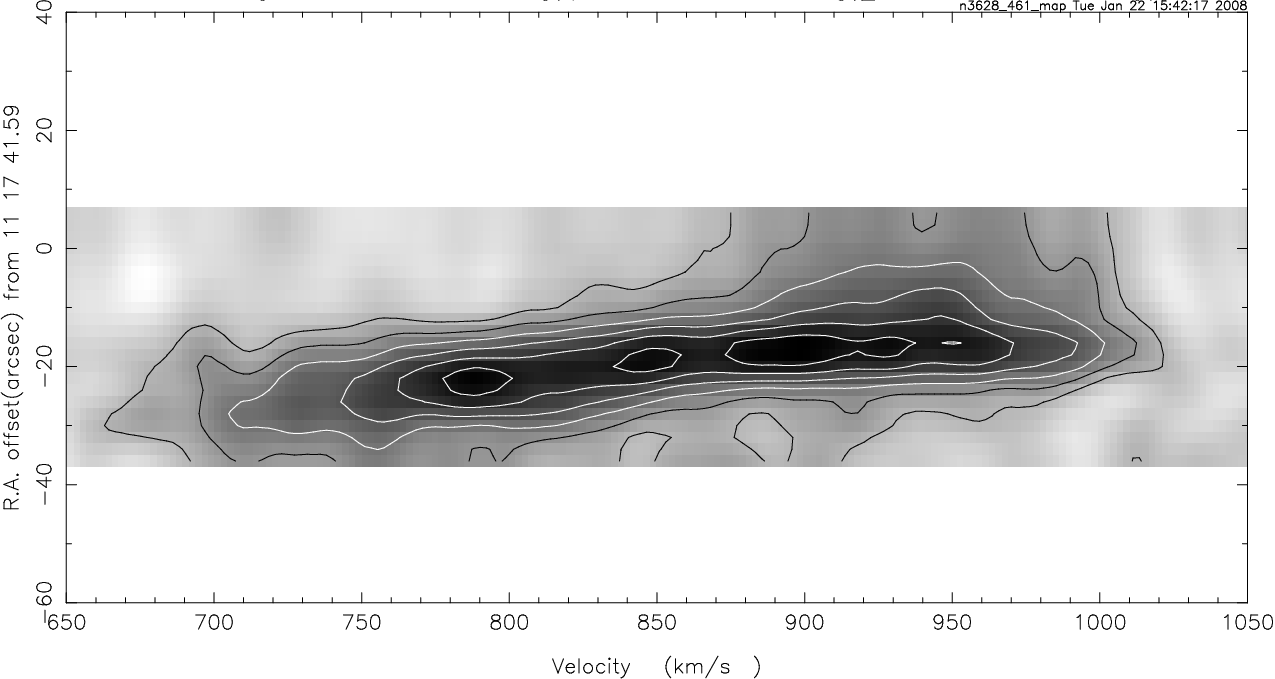}}}
\end{minipage}
\caption[] {Center of NGC~3628. Top: (Left) J=4-3 $^{12}$CO emission
  integrated over the velocity interval 650 - 1050 $\kms$; contour
  step is 40 $\kkms$; (Right) [CI] emission integrated over the same
  velocity interval; contour step is 5 $\kkms$, lowest contour is at 20
  $\kkms$.  Bottom: J=4-3 $^{12}$CO position-velocity map in position
  angle PA = 105$^{o}$ integrated over a strip 20$''$ wide with
  contours in steps of 10 mK; east is at top. }
\label{n3628moremaps}
\end{figure}

Fig.\,\ref{n3628COmaps} shows bright CO emission from the galaxy,
strongly concentrated in the inner $30''$ (i.e. within radii $R\,\leq$
850 pc). The overall extent suggested by our $J$=3-2 $\co$ single-dish
measurements is in excellent agreement with the size shown in the
$J$=1-0 $\co$ array map by Irwin $\&$ Sofue (1996).  The major-axis
velocity diagram shows two kinematic components. A fast-rotating
feature in the center extends over only half this size. It is
superposed on molecular gas in more leisurely rotation extending over
the full area mapped in CO.  The extent of the fast-rotating component
is the same as that of the OH/HI absorption feature measured by
Schmelz et al. (1987a, b).  The velocity gradient of rotation is 360
$\kms$/kpc, corrected for inclination - very similar to that of
NGC~660, but the full rotational width of 305 $\kms$ (again corrected
for inclination) is much less than that of NGC~660.  The molecular gas
distribution in Figs.\,\ref{n3628COmaps} and \ref{n3628moremaps} does
not exhibit an obvious central minimum such as most other galaxies do
but we cannot rule out that the limited spatial resolution and the
steep velocity gradient conspire to hide a small central minimum.  The
IRAM $J$=2-1 $\co$ maps by Reuter et al. (1991) suggest that this is
indeed the case. The fact that the radio core is about half the size
of the extent of the molecular gas in the fast-rotating component
suggests that much of the star-burst takes place at the inside of this
molecular gas concentration.  Throughout the central region, we have
observed emission from neutral carbon. However, the very center shows
an unresolved (i.e. diameter less than 250 pc) [CI] peak
(Fig.\,\ref{n3628moremaps}).

The line ratios in Table 5 suggest that the observed molecular gas is
only modestly optically thick.  Comparison of the ratios in a $21''$
beam with those in a $12''$ beam implies that the molecular gas in the
central region close to the radio core (marking the star-burst) is
denser or warmer than the gas at slightly greater ($R$ = 350 -- 600
pc) distances to the nucleus.  We suspect that the former is the
dominant effect, because almost all the molecular gas in the center of
NGC 3628 is relatively cold: all model fits, including those shown in
Table 6, solidly suggest predominant temperatures $T_{\rm
  kin}\,=\,20-30$ K.  In model 5, most of the gas is both cool and
dense ($n_{\h2}\,=\,10^{4}\,\cc$) but there is a non-negligible mass
of hot ($T_{\rm kin}\,=\,150$ K) and less dense
($n_{\h2}\,=\,10^{3}\,\cc$) gas.  In model 6, all of the gas is cool,
and most of it is rather tenuous ($n_{\h2}\,=\,10^{2}\,\cc$).  In this
model, there is only a small mass of cool gas at higher densities
($n_{\h2}\,=\,10^{4}\,\cc$).  Given the star-burst nature of NGC~3628,
model 5 appears physically more plausible, as is also suggested by the
presence of warm dust in NGC~3628 (Rice et al. 1988).  In terms of
beam-averaged quantities, the difference between the two models is
slight.  In model 5 almost all (95$\%$) of all carbon is in CO.  In
model 6 this fraction is somewhat lower (75$\%$). Both have
beam-averaged $\h2$ column densities
$N_{\h2}\,\approx\,5\times10^{21}\,\cm2$ which is very similar to the
peak HI column density listed by Braun et al. (2007). These column
densities are relatively high but this is not surprising given the
edge-on orientation of the galaxy.  Because of this, we could not
reliably estimate the face-on mass-density.  The central molecular gas
mass is $M_{\h2}\,=\,1.5\times10^{8}\,\Msun$ within $R$ = 0.6 kpc,
independent of the model assumed.  Towards the center of NGC~3628, the
CO-to$\h2$ conversion factor is $X$ = 2.4 $\times$10$^{19}$
$\cm2/\kkms$, i.e. 8 times {\it lower} than the `standard' Solar
Neighborhood value.  A roughly similar value was also estimated by
Irwin $\&$ Sofue (1996). The molecular mass in the central region
calculated by us is typically 2-6$\%$ of the dynamical mass.

\subsection{NGC~4631}

\begin{figure}[]
\unitlength1cm

\begin{minipage}[t]{9.cm}
\resizebox{9.cm}{!}{\rotatebox{270}{\includegraphics*{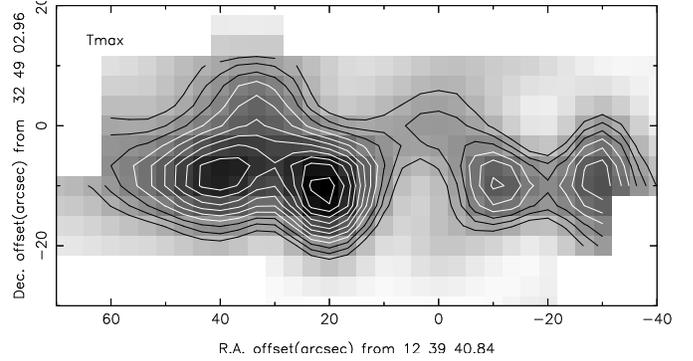}}}
\end{minipage}
\caption[] {Center of NGC~4631. Map of peak J=2-1 $^{12}$CO brightness
  temperature $T_{mb}$. Contour step is 50 mK, lowest contour is at
  300 mK.}
\label{N4631_Peaks}
\end{figure}

The CO maps in Fig.\,\ref{n4631COmaps} are more detailed than those
published earlier by Golla $\&$ Wielebinski (1994) and Dumke et
al. (2001), but similar overall aspect.  The maps published by these
authors show that the most intense CO emission ranges from
$\Delta\alpha$ = +60$'$ to $\Delta\alpha$ = -40$''$ in our map which
just covers this range.  CO intensities are at a relative minimum at
the position of the nucleus of the galaxy, which is located not at the
map (0,0) position, but offsets $\Delta\alpha$ = +12.5$''$,
$\Delta\delta$ = -6$''$.  In the integrated intensity maps, nor in the
major axis-velocity map is there any sign of molecular gas associated
with the nucleus, a conclusion we share with Rand (2001). In addition,
the velocity-integrated CO brightness temperatures are much lower in
NGC~4631 (and NGC~278) than in galaxies with distinct central CO
concentrations such as, for instance, NGC~660 and NGC~3628. There are,
however, several unresolved CO intensity peaks at various distances to
the mid-plane of the galaxy.  These might be thought to reflect spiral
arm tangential locations but this is not very probable because they
are not placed symmetrically with respect to the nucleus. More likely
they are instead giant molecular cloud complexes hosting the burst of
star formation that is taking place in the central region of
NGC~4631. In support of this, we note that they tend to coincide with
radio continuum peaks in the map by Golla (1999), while the
easternmost and brightest CO peak also corresponds to the H$\alpha$
emission region CM~67 (Crillon $\&$ Monnet 1969). The peak CO
brightness temperatures in Fig.\,\ref{N4631_Peaks} are $T_{mb}$ =
0.5-0.8 K. These are almost identical to the brightness temperatures
found for similar CO peaks in NGC~278 and suggest filling factors of
the order of 0.02-0.05 when compared with Galactic GMC's of typical
size 100-150 pc. If we could view NGC~4631 face-on, it would probably
much resemble NGC~278. Contrary to earlier interpretations, we do not
find the observed CO structure of NGC~4631 to suggest the presence of
ring- or bar-like structures in the galaxy.

The line ratios of the diffuse gas in the line of sight through the
center of NGC~4631 (Table 5) do not allow much choice in model
parameters. Twenty per cent of the gas mass sampled in this direction
is warm ($T_{kin}\,\approx\,150$ K) and moderately dense
($n_{\h2}\,\approx\,1000\,\cc$) but most (80$\%$) of it is cold
($T_{kin}\,\approx\,10$ K) and relatively tenuous
($n_{\h2}\,\approx\,300\,\cc$). In this particular direction which
does not sample an active center nor an actively star forming region,
beam-averaged $\h2$ column densities are not particularly high
(cf. Table 7) and most (70$\%$) of the carbon appears to be in atomic
form. Towards the center of NGC~4631 we find a CO-to$\h2$ conversion
factor $X$ = 3.1 $\times$10$^{19}$ $\cm2/\kkms$, i.e. 6 times {\it
  lower} than the `standard' Solar Neighborhood value. Paglione et
al. (2001) estimated an $X$-factor only slightly higher than this
towards the center of NGC~4631. Thus, the implied molecular gas mass
in NGC~4631 is much lower, and a much smaller fraction of the
dynamical mass, than use of the `standard' conversion factor would
suggest.

\subsection{NGC~4666}

\begin{figure}[]
\unitlength1cm

\begin{minipage}[t]{5.cm}
\resizebox{5.cm}{!}{\rotatebox{270}{\includegraphics*{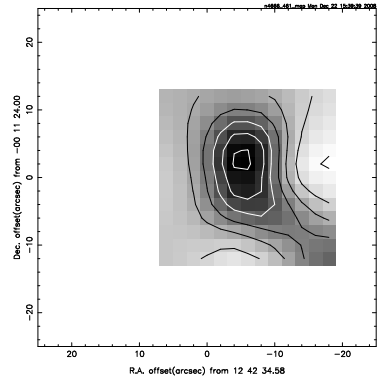}}}
\end{minipage}
\caption[] {Center of NGC~4666. Map of velocity-integrated J=4-3 $^{12}$CO brightness
  temperature $T_{mb}$d$V$. Contour step is 1.5 $\kkms$, lowest contour is at
  3 $kkms$}
\label{n4666more}
\end{figure}

The maps in Fig.\,\ref{n4666COmaps} show CO emission to extend over
$120''$ (13 kpc) in the plane of the galaxy. This extent only
marginally exceeds that of the CO emission mapped at high resolution
by Walter et al. 2004 (their Fig. 8). The $\co$ maps show a strong
central peak, which we also observed in the $J$=4-3 transition
(Fig.\,\ref{n4666more}).  Although the slightly asymmetrical CO
distribution in our $J$=3--2 $\co$ position-velocity map suggests a
small offset from the nucleus, this central CO peak is in fact
coincident with the kinematic center of the galaxy derived from the HI
emission by Walter et al. (2004). Nevertheless, their CO array maps
show that the central molecular gas maximum seen in our maps in the
lower $J$ transitions is, in fact, a blend of a compact CO maximum
centered on the nucleus and star-forming molecular clouds in the disk
close to the nucleus. The compact appearance of the central
condensation in the $J$=4-3 $\co$ map (Fig.\,\ref{n4666more}) implies
that it is significantly denser than the gas farther from the nucleus.

The inner parts of NGC~4666 rotate fast, with an inclination-corrected
velocity span of 360 $\kms$ over the central arcminute. Unlike most of
the other galaxies mapped by us, the inner parts of NGC~4666 do not
quite follow a solid-body rotation curve. 

We note that the $\co$/$\13co$ ratios of NGC~4666 are relatively low
and similar to those of NGC~278. As is the case in that galaxy, such
values suggest the relatively high CO optical depths characteristic of
star-forming Giant Molecular Clouds in the Milky Way disk. Our model
analysis in Tables 6 and 7 yields roughly equal masses of modestly
dense ($n_{\h2}\approx10^{3} \cc$) and hot ($T_{kin}\approx100$ K) gas
and dense ($n_{\h2}\approx10^{4} \cc$) and cold ($T_{kin}\approx10$ K)
gas.  There are roughly equal masses of neutral carbon and carbon
monoxide. Molecular hydrogen column densities, central mass, and
face-on mass density do not stand out with respect to the other
galaxies discussed in this paper.  The central CO-to-$\h2$ conversion
factor implied by Tables 4 and 7 is only $X$ = 1.4 $\times$10$^{19}$
$\cm2/\kkms$, a factor 14 lower than the Solar Neighborhood
standard. The central mass is about 0.5 per cent of the dynamical
mass.

\subsection{Column densities and masses}

For the galaxies in this paper, we find $\h2$ column densities
averaged over a $21''$ beam of roughly 0.5-5 $\times\,10^{21}\,\cm2$
(corresponding to {\it total hydrogen} columns of 1-10
$\times\,10^{21}\,\cm2$). Thus, these galaxies cover a somewhat
greater range than the 2-6 $\times\,10^{21}\,\cm2$ that we found for
more active galaxies in our previous paper (Israel, 2009) and they are
mostly somewhat lower.  These total columns also include the ambient
contribution by neutral atomic hydrogen which we have so far
neglected.  Previously, we have found that HI {\it absorption} column
densities actually measured usually exceed the {\it beam-averaged}
$\h2$ column densities we derive.  This is also the case for the only
two galaxies with HI absorption results in the literature, NGC~660
(Baan et al. 1992) and NGC~3628 (Schmelz et al. (1987a), with typical
HI absorption column densities $N(HI)\,\approx\,10^{22}\,\cm2$,
i.e. 4-7 the value listed in Table 7 for NGC~660, and twice the value
given for NGC~3628.  However, the effective absorbing area is smaller
by at least an order of magnitude than the large $21''$ beam over
which we have averaged our $\h2$ in Table\,\ref{galresult}.  More
importantly, in these highly tilted galaxies much of the HI seen in
absorption must reside in the disk, and the actual HI content of the
central region is unknown (see also the discussion by Baan et
al. 1992).  Although we have no direct knowledge of central HI column
densities in the present sample, we note that most (if not all)
galaxies with strong central molecular gas concentrations exhibit a
pronounced lack of HI emission at their centers. In the six more or
less face-on galaxies previously studied by us, we found a mean value
$N(HI)\,=\,0.6\pm0.3\,\times10^{21}\,\cm2$, and there is no reason to
assume that the galaxies discussed here are very different.  Indeed,
in the HI {\it channel maps} of NGC~4666, there is a clear lack of
neutral hydrogen at the position of the nuclear CO peak (Walter et
al. 2004, their Figure 9).  From the previous results, we estimate
that the HI contribution to the total hydrogen column densities
derived by us is small, typically 10 per cent uncertain by a factor
two either way.  However, the one face-on galaxy discussed here,
NGC~278, clearly lacks a central molecular concentration and here the
HI contribution may be significant.

The $\h2$ masses derived are again much lower than those calculated
previously by authors who assumed the `standard' CO-to-$\h2$
conversion factor $X$ to apply.  If we were to have underestimated
the HI contribution, or to have overestimated the CO column density,
these $\h2$ masses would be even lower.  With the results presented for
the galaxy centers in this paper, we find $X$ = 0.1-0.3
$\times\,10^{20}\,\cm2/(\kkms)$ which is a factor of 7-20 lower than
the `standard' Milky Way $X$ factors (see also Strong et al. 2004),
but quite similar to those found in other galaxy centers (Israel 2009,
 Israel $\&$ Baas 2003 and references therein).  We also note that at
least for NGC~3628 our present result is practically identical to the
low value ($X$ = 0.2-0.7 $\times\,10^{20}\,\cm2/(\kkms)$ found
independently by Irwin $\&$ Sofue (1996). With the low $X$ values
presented here, the molecular gas does not constitute more than
typically a few per cent of the total (dynamical) mass in the same
volume.

\subsection{Temperature and excitation}

Although a classical UV-illuminated photon-dominated region (PDR)
associated with a star-forming region may reach quite high kinetic
temperatures, the mass of gas at such elevated temperatures is
rather small and limited to the immediate vicinity of the stellar
heating source.  Even in intense star-forming regions globally
averaged temperatures are unlikely to be in excess of 20-30 K.

When we look at Table\,\ref{galmodel} we see that only towards the
center of the fully edge-on galaxy NGC~3628 the molecular line
emission may arise from a gas purely dominated by {\it star-bursts}.
In the other four galaxy centers, a 10-30 K molecular gas component is
always accompanied by a much hotter (usually relatively tenuous)
component of temperature 100-150 K. In NGC~278 and NGC~660 this hot
component contains, in effect, most of the molecular gas mass, in
NGC~4666 it is about half, and only in the quiescent center of
NGC~4631 it is a small ($20\%$) fraction. Although the galaxies
discussed here all have {\it star-burst-dominated} central regions,
they are thus mostly too hot to be excited only by the UV photons from
luminous stars.  We found a similar result in the preceding paper on
five AGN/star-burst galaxies (Israel 2009). There, as is the case here,
central star-bursts were seen to produce extended outflows glowing in
soft-X-ray emission.  Also the galaxies included in this paper support
the suggestion that the hot molecular gas component is efficiently
heated by the X-rays from these flows (see Meijerink et al. 2006,
2007) rather than by the UV photons from the star-burst.

\subsection{Comparison with AGN central CO emission}

A brief comparison with the galaxies containing a (modest) AGN,
discussed in our previous paper (Israel, 2009) shows a number of
resemblances and differences. In galaxy centers of either type, we
need to fit the observed CO lines with at least two different
molecular gas density/temperature components: a hot and tenuous one,
and a cooler and dense one. However, we find that the highest
densities occur in galaxy centers that have both a star-burst and an
AGN. The size of the central molecular gas concentration is similar
whether or not an AGN is present, but its mass and column density are
on average lower in the star-burst-only than in the combined
star-burst/AGN galaxies. CO emission is also weaker in star-burst-only
galaxies, which also tend to have lower $\co$/$\13co$ ratios, hence
higher average CO optical depths. The {\it weakest} CO galaxies in the
sample (NGC~278 and NGC~4631) are pure star-burst galaxies {\it
  without} a significant central CO concentration.  Rather, these
galaxies confine their star-burst activity to the disk spiral arms.
In galaxies of either type, the CO-to-$\h2$ conversion factors $X$
appropriate to the central region are low - there is no systematic
difference. In terms of excitation, both galaxy center types require
another mechanism to operate in addition to the UV excitation from
PDRs. X-rays from diffuse gas heated by star-burst-driven winds may be
the answer.  In any case and at least on scales of a few hundred
parsecs or more, (modest) AGNs do not seem to leave an unambiguous
imprint on the centrally concentrated molecular gas.

\section{Conclusions}

\begin{enumerate}
\item We have observed the centers of five galaxies undergoing
  significant star-burst activity in the first four transitions of
  $\co$ and the first three transitions of $\13co$, and three of them
  also in the lower transition of [CI].
\item All galaxies were mapped over at least $1'\times1'$ in the
  $J$=2-1 and $J$=3-2 transitions and over a smaller area in the
  $J$=4-3 transition of $\co$.
\item The central line ratios, reduced to a standard $21''$ beam. vary
  significantly in the sample galaxies. In general, relatively strong
  $J$=4-3 line emission indicates elevated densities and perhaps
  temperatures for the molecular gas.
\item $\co$ optical depths appear not to be very high. NGC~660,
  NGC~3628 and NGC~4631 have $\co/\13co$ ratios in the range of
  11-16.  NGC~278 and NGC~4666 have lower isotopic ratios in the range
  6-8 suggesting the somewhat higher optical depths reminiscent of
  extended molecular clouds in the Solar Neighborhood
\item In NGC~660, NGC~3628, and NGC~4666 molecular gas is present in
  the form of clearly identifiable central concentrations of outer
  radius $R=0.6-1.5$ kpc. Such central concentrations are lacking in
  NGC~278 and NGC~4631. The central concentrations in NGC~660 and
  NGC~3628 exhibit a local minimum at the nucleus itself.
\item In the central regions of NGC~660, NGC~3628, and NGC~4666, we
  find moderate $\h2$ masses of $0.6-1.5\times10^{8}\,\Msun$. These
  masses are typically about one per cent of the dynamical mass in the
  same region.  The CO-to-$\h2$ conversion factors $X$ implied by
  these masses and the observed $J$=1--0 $\co$ line intensities is
  typically an order of magnitude lower than the `standard' Solar
  Neighborhood $X$-factor.
\item The molecular gas of all observed galaxies may at least partly
  be excited by UV-photons from the extended circumnuclear star-burst,
  but in all galaxies (except perhaps NGC~3628) a significant fraction
  of the molecular gas must be excited in other ways such as the X-rays
  seen to emanate from massive wind-driven outflows. 
\end{enumerate}

\acknowledgements{I am much indebted to the late Fred Baas, who died
  suddenly and far too early in 2001, for carefully planning and
  performing many of the observations described in this paper.  I also
  thank the various JCMT observers who made additional observations
  in queue mode.  Finally, it is always a pleasure to thank the
  personnel of both the JCMT and the IRAM 30m telescope for their able
  and generous support.  }

\end{document}